\documentclass[]{emulateapj}
\usepackage{natbib,amsmath}

\shorttitle{Probing the IGM/Galaxy Connection V: 
Origin of \lya\ and \ovi\ Absorption}
\shortauthors{Prochaska et al.}

\begin{document}

\def \hkpc {$h_{72}^{-1}$\,kpc}
\def \hvkpc {$h_{70}^{-1}$\,kpc}
\def \msun {M_\odot}
\def \rcgm {$r_{\rm CGM}$}
\def \mrcgm {r_{\rm CGM}}
\def \mncgm {N_{\rm H, CGM}}
\def \lowz {low-$z$}
\def \mnh {n_{\rm H}}
\def \mlolx {\ell_{\rm Ly\alpha}(X)}
\def \lolx {$\ell_{\rm Ly\alpha}(X)$}
\def \lolz {$\ell_{\rm Ly\alpha}(z)$}
\def \loox {$\ell_{\rm OVI}(X)$}
\def \mloox {\ell_{\rm OVI}(X)}
\def \lox {$\ell(X)$}
\def \mlox {\ell(X)}
\def \loz {$\ell(z)$}
\def \looz {$\ell_{\rm OVI}(z)$}
\def\subls{sub-$L^*$} 
\def\lstar{$L^*$} 
\def\wlya{$W^{\rm Ly\alpha}$}
\def\mwlya{W^{\rm Ly\alpha}}
\def\mwovi{W^{\rm 1031}}
\def\wovi{$W^{\rm 1031}$}
\def\mnovi{N_{\rm OVI}}
\def\novi{$N_{\rm OVI}$}
\def\nfields{$\approx 20$}
\def\tonso{TonS~180}
\def\tonst{TonS~210}
\def\mug{(u-g)}
\def\ug{$(u-g)$}
\def\rcgm{$r_{\rm CGM}$}
\def\mrcgm{r_{\rm CGM}}
\def\mrvir{r_{\rm vir}}
\def\rvir{$r_{\rm vir}$}
\def\npub{13}
\def\nanly{14}
\def\vbestmfp{48.2 \pm 1.9}
\def\vslopemfp{-37 \pm 4.7}
\def\bestmfp{\lambda_0}
\def\slopemfp{b_\lambda}
\def\omegam{\Omega_{\rm m}}
\def\mnull{\nu_{\rm 912}}
\def\nnull{$\nu_{\rm 912}$}
\def\hub{h_{72}^{-1}}
\def\umfp{{\hub \, \rm Mpc}}
\def\lbr{$\lambda_{\rm r}$}
\def\mlbr{\lambda_{\rm r}}
\def\intl{\int\limits}
\def\kll{$\kappa_{\rm LL}$}
\def\mkll{\kappa_{\rm LL}}
\def\kconst{$\kappa_{\mzq}$}
\def\mkconst{\kappa_{\mzq}}
\def\mztkll{{\tilde\kappa}_{912}}
\def\zq{$z_q$}
\def\mzq{z_q}
\def\maxoff{0.4}
\def\clls{3.50 \pm 0.12}
\def\alls{1.89 \pm 0.39}
\def\blls{-1.0 \pm 0.3}
\def\cmma{\;\;\; ,}
\def\perd{\;\;\; .}
\def\ltk{\left [ \,}
\def\ltp{\left ( \,}
\def\ltb{\left \{ \,}
\def\rtk{\, \right  ] }
\def\rtp{\, \right  ) }
\def\rtb{\, \right \} }
\def\sci#1{{\; \times \; 10^{#1}}}
\def \rAA {\rm \AA}
\def \zem {$z_{\rm em}$}
\def \mzem {z_{\rm em}}
\def \mzlls {z_{\rm LLS}}
\def \zlls {$z_{\rm LLS}$}
\def \mzend {z_{\rm end}}
\def \zend {$z_{\rm end}$}
\def \zstrto {$z_{\rm start}^{\rm S/N=1}$}
\def \mzstrto {z_{\rm start}^{\rm S/N=1}}
\def \zstrt {$z_{\rm start}$}
\def \mzstrt {z_{\rm start}}
\def\smm{\sum\limits}
\def \lll  {$\lambda_{\rm 912}$}
\def \mlll  {\lambda_{\rm 912}}
\def \mtll  {\tau_{\rm 912}}
\def \tll  {$\tau_{\rm 912}$}
\def \avtll  {\tilde{\mtll}}
\def \tigm  {$\tau_{\rm IGM}$}
\def \mtigm  {\tau_{\rm IGM}}
\def \cmm  {cm$^{-2}$}
\def \cmmm {cm$^{-3}$}
\def \kms  {km~s$^{-1}$}
\def \mkms  {{\rm km~s^{-1}}}
\def \lyaf {Ly$\alpha$ forest}
\def \Lya  {Ly$\alpha$}
\def \ovi  {\ion{O}{6}}
\def \mlya  {\rm Ly\alpha}
\def \lya  {Ly$\alpha$}
\def \mlya  {Ly\alpha}
\def \Lyb  {Ly$\beta$}
\def \lyb  {Ly$\beta$}
\def \Lyg  {Ly$\gamma$}
\def \lyg  {Ly$\gamma$}
\def \lyd  {Ly$\delta$}
\def \ly5  {Ly-5}
\def \ly6  {Ly-6}
\def \ly7  {Ly-7}
\def \nhi  {$N_{\rm HI}$}
\def \mnhi  {N_{\rm HI}}
\def \lnhi {$\log N_{HI}$}
\def \mlnhi {\log N_{HI}}
\def \etal {\textit{et al.}}
\def \ob {$\Omega_b$}
\def \obh {$\Omega_bh^{-2}$}
\def \om {$\Omega_m$}
\def \ol {$\Omega_{\Lambda}$}
\def \gz {$g(z)$}
\def \mgz {g(z)}
\def \lyaf {Lyman--$\alpha$ forest}
\def \fnhi {$f(\mnhi,z)$}
\def \mfnhi {f(\mnhi,z)}
\def \mfp {$\lambda_{\rm mfp}^{912}$}
\def \mmfp {\lambda_{\rm mfp}^{912}}
\def \btlls {$\beta_{\rm LLS}$}
\def \mbtlls {\beta_{\rm LLS}}
\def \teff {$\tau_{\rm eff,LL}$}
\def \mteff {\tau_{\rm eff,LL}}
\def \O {${\mathcal O}(N,X)$}
\newcommand{\cm}[1]{\, {\rm cm^{#1}}}
\def \zem {$z_{em}$}
\def \snrlim {SNR$_{lim}$}

\title{Probing the IGM/Galaxy Connection V: On the Origin of
  \lya\ and \ovi\ Absorption at $z<0.2$}

\author{
J. Xavier Prochaska\altaffilmark{1}, 
B. Weiner\altaffilmark{2}, 
H.-W. Chen\altaffilmark{3}, 
J. Mulchaey\altaffilmark{4}, 
K. Cooksey\altaffilmark{5}, 
}
\altaffiltext{1}{Department of Astronomy and Astrophysics \& UCO/Lick
  Observatory, University of California, 1156 High Street, Santa Cruz,
  CA 95064; xavier@ucolick.org}
\altaffiltext{2}{Steward Observatory, University of Arizona, 933
  N. Cherry Ave., Tucson, AZ 85721; bjw@as.arizona.edu}
\altaffiltext{3}{Department of Astronomy; University of Chicago; 5640
  S. Ellis Ave., Chicago, IL 60637; hchen@oddjob.uchicago.edu} 
\altaffiltext{4}{Carnegie Observatories; 813 Santa Barbara St.,
  Pasadena, CA 91101; mulchaey@obs.carnegiescience.edu}
\altaffiltext{5}{NSF Astronomy \& Astrophysics Postdoctoral Fellow, MIT
  Kavli Institute for Astrophysics \& Space Research, 77 Massachusetts
  Avenue, 37-611, Cambridge, MA 02139, USA; kcooksey@space.mit.edu}

\begin{abstract}
We analyze the association of galaxies to \lya\ and \ovi\ absorption,
the most commonly detected transitions in the \lowz\ intergalactic
medium (IGM), in the fields of \nanly\ quasars with $\mzem =
0.06$--$0.57$.  Confirming previous studies, we observe a high covering
fraction for \lya\ absorption 
to impact parameter $\rho = 300$\,\hkpc:
33/37 of our $L > 0.01L^*$ galaxies show \lya\ equivalent width
$\mwlya \ge 50$\,m\AA. 
Galaxies of all luminosity $L > 0.01L^*$ and spectral type are
surrounded by a diffuse and ionized circumgalactic medium (CGM), whose
baryonic mass is estimated at $\sim 10^{10.5 \pm 0.3} \msun$ for a
constant $N_{\rm H}$.  The virialized
halos and extended CGM of present-day galaxies are responsible for most
strong \lya\ absorbers ($\mwlya > 300$\,m\AA) but cannot
reproduce the majority of observed lines in the \lya\ forest.  We
conclude that the majority of \lya\ absorption with $\mwlya =
30$--$300$\,m\AA\ occurs in the cosmic
web predicted by cosmological simulations and estimate a characteristic
width for these filaments of $\approx 400$\,\hkpc.
Regarding \ovi, we observe a near unity covering fraction to $\rho
= 200$\,\hkpc\ for $L>0.1L^*$ galaxies and to $\rho = 300$\,\hkpc\
for \subls\ ($0.1 L^* < L < L^*$) galaxies.  Similar to our \lya\ results, 
stronger \ovi\ systems ($\mwovi > 70$\,m\AA) arise in the virialized
halos of $L>0.1L^*$ galaxies.  Unlike \lya, the weaker \ovi\
systems ($\mwovi \approx 30$\,m\AA) arise in the extended CGM of
\subls\ galaxies.  The majority of \ovi\
gas observed in the \lowz\ IGM is associated with a 
diffuse medium surrounding individual galaxies with $L \approx 0.3
L^*$, and rarely originates in the so-called warm-hot IGM (WHIM)
predicted by cosmological simulations.
\end{abstract}


\section{Introduction}

Ultraviolet absorption-line spectroscopy remains the most
efficient and sensitive means of studying the diffuse gas that
permeates the universe.  This gas, commonly referred to as the
intergalactic medium (the IGM) or \lya\ forest, is the dominant reservoir
of baryons in the universe at all epochs \citep{pt09}. 
It provides fresh fuel for star-formation and collects the radiation
and metals produced by galaxies and active galactic nuclei. 
Furthermore, the IGM is expected to trace the
dark matter distribution of large-scale structure, offering unique
constraints on our cosmological paradigm
\citep[e.g.][]{mco+96,rau98,msb+06,vbh09}.

Analysis of quasar spectra acquired with UV-sensitive spectrometers on
the {\it Hubble Space Telescope} ({\it HST}) and 
{\it Far-Ultraviolet Spectroscopic Explorer (\it FUSE)} have
surveyed the redshift distribution, metal-enrichment, ionization state, and temperature 
of the IGM in the $z<1$ universe
\citep[e.g.][]{wjl+98,dhk+99,pwc+06,tc08a,tripp08,ds08,ctp+10}.
These studies have demonstrated, similar to the high $z$ IGM,
that the
gas collects in a set of relatively discrete `lines' in redshift, has
a high ionization fraction, and is often enriched with heavy metals (e.g.\ C, O).
The \lya\ forest is significantly sparser at $z<1$, however, because
of the steady expansion of the universe.  
The cosmic abundance of heavy metals also appears to have evolved
\citep[e.g.][]{pks0405_uv,ds08,ctp+10}, although the scatter
for individual regions is large.

Over the past decade, extra attention has been placed
on the search for a
warm-hot phase of the IGM $(T \gtrsim 10^6$K) that is predicted by
cosmological simulations to be a major baryonic reservoir in the
low $z$ universe \citep{co99,daveetal01}.
In the UV, the search for this so-called WHIM (warm-hot intergalactic
medium) includes surveys for the \ion{O}{6} doublet
\citep[e.g.][]{tripp08,tc08a} 
searches for thermally broadened \lya\ lines
\citep[e.g.][]{lsr+07}
and, most recently, surveys for \ion{Ne}{8} absorption
\citep{savage05,nws09}.
Owing to its greater ease of detection, the \ovi\ doublet has been the most
frequently studied transition to date.
Focused surveys have characterized its incidence as a
function of equivalent width and column density
\citep{tripp08,tc08a,ds08,wakker09}, and have permitted
statistical comparison with predictions from cosmological simulations
\citep[e.g.][]{cen09,od09}.

Although UV spectral analysis provides robust constraints
on the nature and distribution of diffuse gas, considerable debate
remains over the origin and cosmological relevance of the observed
absorption systems.  For example, hot and diffuse gas is predicted
(and observed) to
exist in the outer halos of individual galaxies
\citep{spitzer56,bs69,mm96,mb04,sws+03}, the
intragroup medium \citep[e.g.][]{mulchaey96,fcw08}, and also the filamentary
structures of the IGM \citep[e.g.][]{co99}.  The incidence of
\ion{O}{6}, therefore, is likely a sensitive function of how metals are
dispersed on galactic and intergalactic scales and also the physical
conditions within these various environments \citep[e.g.][]{od09}. 
Not surprisingly, groups have drawn competing conclusions on whether
the \ovi\ absorbers primarily trace the low-density WHIM or a photoionized,
cooler phase.
Similarly, multiple scenarios may explain the incidence and properties
of the \lya\ `clouds' along single IGM sightlines.  While ram-pressure
stripping, tidal stripping, and galactic-scale winds are all believed
to play a role in transporting gas from galaxies to the IGM, the
timing and relative importance of each of these remain open issues.
In short, there remain several fundamental questions on the origin
and nature of the two most common transitions of the IGM.

Researchers have long recognized that additional insight into the IGM may be
gained by surveying the fields surrounding absorption-line systems
for galaxies and their large-scale structures.  Early efforts focused on
\lya\ absorption and the relation of HI gas to galaxies and
large-scale structures \citep{sfy+93,mwd+93,lbt+95}. 
A key results was that sightlines with small impact parameters to
galaxies ($\rho \lesssim 250$\,\hkpc) had a high incidence of moderate
to strong \lya\ absorption \citep{lbt+95,clw+98,tripp+98,chenetal01,bowen+02,mj06}.
This indicates galaxies are surrounded by a diffuse, highly
ionized medium that gives rise to significant \lya\ absorption
\citep[see also][]{wakker09}.
These studies were followed by two-point cross-correlation analysis of
galaxies to absorbers on $\approx 1$\,Mpc scales
\citep{cpw+05,wmj+07,cm09} and have recently been 
extended to $z \sim 1$ \citep{shone+10}. 
Their results reveal significant clustering between
galaxies and strong \lya\ systems (absorbers with HI column density,
$\mnhi > 10^{15} \cm{-2}$) that implies a causal connection.  In
contrast, they measure a very
weak or no clustering signal for low-\nhi\ absorption
systems which implies the two phenomena trace
different structures in the universe. 
While previous work has offered valuable insight into the nature of
the \lya\ forest, it is limited by sample variance both in terms of
the number of absorbers, galaxies, and fields surveyed.  

Several projects have now also considered the galaxy/absorber connection
for the metal-enriched IGM, e.g.\ 
\ion{Mg}{2}, \ion{C}{4}, and \ion{O}{6} selected systems. 
For the first, 
observers have commonly associated strong \ion{Mg}{2} absorption
with individual galaxies at impact parameters $\rho < 100$\,kpc,
by searching for galaxies at the redshifts of known \ion{Mg}{2}
systems \citep[e.g.][]{bergeron86,bb91}.
Indeed, galaxies at \lowz\ show modest to high covering
fractions of cool
gas ($T \sim 10^4$\,K) traced by \ion{Mg}{2} absorption
\citep{bc09,ct08,chg+10}.
This has led to the association of metal-enriched, cool IGM gas with the
`halos' of individual galaxies.  
Galaxies also exhibit a high covering fraction to \ion{C}{4}
absorption for impact parameters $\rho \lesssim 200$\,kpc
\citep{clw01}.  Although the origin of this highly ionized gas is not
well established, these results likely require a multi-phase medium in
the virialized halos of modern galaxies \citep{mm96}.

The \ovi\ doublet has long been recognized to trace a highly ionized,
warm/hot phase of the Galaxy.  The original detections associated
\ovi\ gas with a Local Bubble surrounding the Sun
\citep{york74,sl06}.  A statistical analysis of \ovi\ absorption
towards extragalactic sources with {\it FUSE} suggests a thick layer
of highly ionized gas with a scale height of $\approx 3$\,kpc
\citep{ssj+00}.  Lastly, studies of \ovi\ absorption toward
extragalactic sources reveal a high covering fraction from an inferred
halo of collisionally ionized gas at $T \approx 10^5$\,K
\citep{savage79,ssw+03,sws+06}.  Similar halos of
\ovi-bearing gas have been inferred for external galaxies through
\ovi\ emission maps \citep{bma+05}. 
It is evident that some fraction
of \ovi\ observed in the IGM must arise from galactic halos.

Connecting \ovi\ gas of the IGM to galaxies and the structures they
reside within has been the focus of several recent studies.
In \cite{pwc+06}, we examined the field surrounding the multiple \ion{O}{6}
systems identified along the sightline to PKS0405--123.  While the
strong absorption at $z=0.167$ was notable
for its association with two galaxies at small impact parameter ($\rho \sim
100$\,kpc), additional \ovi\ systems showed no obvious galactic counterparts
and/or galaxies only at large impact parameter. 
A similar diversity of galactic environment has been observed for
other sightlines \citep{tsg+05,sts+04,cpc+08,cm09}.
\cite{stockeetal06} performed the first multi-sightline analysis on
the galaxy/absorber connection for \ovi.  For the 9 \ovi\ systems in
their fields complete to 0.1~\lstar\ galaxies, they estimated a 
median distance to an $L \ge 0.1 L^*$ galaxy of $\approx 180$\,\hvkpc.
Furthermore, they identified an \lstar\ galaxy within 800\,\hvkpc\ for
essentially each of the 23 \ovi\ absorbers in their survey.  They
inferred, based on these results, that \ovi\ gas does not occur in
galaxy voids. 
They further hypothesized that galaxies with $L \le 0.1 L^*$ galaxies
may be most responsible for \ovi\ absorption.
Similar results were found by \cite{wakker09} for a small sample of
$z\approx 0$ galaxies.

\begin{deluxetable*}{cccccccccc}
\tablewidth{0pc}
\tablecaption{LCO/WFCCD Fields Analyzed\label{tab:fields2}}
\tabletypesize{\scriptsize}
\tablehead{\colhead{Quasar} & \colhead{RA} & \colhead{DEC} & 
\colhead{$z_{\rm em}$} & \colhead{{\it HST} UV Spectroscopic Datasets} & 
\colhead{{\it FUSE}$^a$} & \colhead{\lya\ Ref.} & \colhead{\ovi\ Ref.}
& \colhead{$\mathcal{N}_{\rm spec}^b$} \\
& (J2000) & (J2000)
}
\startdata
Q0026+1259 & 00:29:13.8 & +13:16:04       & 0.142 & GHRS/(G270M)& 20&&9& 60\\
\tonso & 00:57:20.0 & --22:22:56       & 0.062 & STIS/(G140M,G230MB)&132&5&4&  7\\
PKS0312-77 & 03:11:55.2 & --76:51:51       & 0.223 & STIS/(E140M)&&3&1,3& 56\\
PKS0405-123 & 04:07:48.4 & --12:11:37       & 0.573 & STIS/(E140M,G230M); GHRS/(G160M,G200M)& 71&3,4,9&1,3,4,9&565\\
PG1004+130 & 10:07:26.1 & +12:48:56       & 0.240 & STIS/(G140M)& 85&9&9& 61\\
HE1029-140 & 10:31:54.3 & --14:16:51       & 0.086 & STIS/(G140M)&&5&&  8\\
PG1116+215 & 11:19:08.70 & +21:19:18.      & 0.176 & STIS/(G140M,E140M,E230M); GHRS/(G140L)& 76&1,3,4,5,9&1,2,3,4& 74\\
PG1211+143 & 12:14:17.7 & +14:03:13.      & 0.081 & STIS/(G140M,E140M); GHRS/(G140L,G270M)& 52&3,4,5&3,4& 25\\
PG1216+069 & 12:19:20.9 & +06:38:38       & 0.331 & STIS/(E140M); GHRS/(G140L)& 13&1,3,6,9&1,2,3,9&101\\
3C273 & 12:29:6.70 & +02:03:9.0      & 0.158 & STIS/(E140M); GHRS/(FG130,FG190,G160M)& 42&1,3,4,9&1,2,3,4& 32\\
PKS1302-102 & 13:05:33.0 & --10:33:19       & 0.286 & STIS/(E140M)&140&3,4,9&1,3,4,9& 63\\
MRK1383 & 14:29:06.4 & +01:17:06.0     & 0.086 & STIS/(E140M,G140M)& 64&3,4,5&4,9&  5\\
FJ2155-0922 & 21:55:01.5 & --09:22:25.0     & 0.192 &
STIS/(E140M,G230MB)& 46&1,3,4,9&1--4,9&105\\
PKS2155-304 & 21:58:51.8 & --30:13:30.0     & 0.116 & STIS/(E140M); GHRS/(G160M,ECH-B,G140L)&123&3,4,9&3,4& 43\\
\enddata
\tablenotetext{a}{Total integration time in ks.}
\tablenotetext{b}{Number of spectroscopically determined galaxy redshifts for objects with $0.005 < z < z_{\rm em}$.}
\tablerefs{
1: \cite{tripp08};
2: \cite{tc08a};
3: \cite{ds08};
4: \cite{dsr+06};
5: \cite{pss04};
6: \cite{cm09};
9: This~paper.
}
\end{deluxetable*}

Recently, we published the results of a galaxy survey performed with
the WFCCD spectrometer on the 100$''$ Dupont telescope at Las Campanas
Observatory (LCO) in the fields surrounding 20 UV-bright quasars at
$z<1$ \citep{ovi_paper4}.  The principal motivation for this survey was to assess the
galaxy/absorber connection for \ovi\ gas at $z \le 0.2$.  In this
paper, we present the first scientific results from our complete survey. 
Previous papers studied the galaxy/IGM connection in a few, individual
fields \citep{pks0405_uv,cpc+08,cpw+05,lpk+09}.
A future paper will study the cross-correlation function between 
galaxies and \lya/\ovi\ absorbers.

This paper is organized as follows.  In $\S$~\ref{sec:data}, we
summarize the datasets that comprise the galaxy surveys \citep{ovi_paper4}
and the characterization of the IGM along the sightlines.  The primary
results of our analysis of galaxy/absorber association for \lya\ and
\ovi\ are given in $\S$~\ref{sec:results}.  We discuss the
implications for the \lowz\ IGM of these results and their context 
with respect to previous work in $\S$~\ref{sec:discussion} and we
conclude with a summary in $\S$~\ref{sec:summary}.
Unless otherwise specified, we adopt a $\Omega_\Lambda=0.72,
\Omega_m=0.28, H_0=72 \rm \, km \, s^{-1} \, Mpc^{-1}$ cosmology
\citep{wmap05}.  Furthermore, all distances are quoted in proper units
unless otherwise noted.

\section{Datasets }
\label{sec:data}

\subsection{The LOC/WFCCD Galaxy Survey}
\label{sec:wfccd}

We have recently published a survey of galaxies in the fields of
20~quasars with bright UV fluxes (Paper~IV), the majority of which have UV spectral
datasets that yield precise measurements of \lya\ and/or \ovi\
absorption (Table~\ref{tab:fields2}).  
The galaxy survey was performed with
the WFCCD spectrometer on the 100$''$ Dupont telescope and was
designed to achieve a high level of completeness to $R \le 19.5$\,mag
galaxies in an $\approx 20' \times 20'$ field centered on each quasar.
These survey parameters were chosen to recover dwarf galaxies ($L< 0.1 L^*$)
to at least $z=0.1$ and to span a $\approx 1$\,Mpc impact parameter at
$z \approx 0.1$.  With this experimental design, we aimed to study the galaxies and
their structures (e.g.\ groups) associated with IGM absorption.
In several analyses (where noted), we also include the galaxy survey in the field of
PKS0405--123 published by \cite{willigeretal06}.

The fields were selected on the
basis of the quasar redshift (higher $z$ was preferred) and UV
flux, with some preference given to fields with existing UV
spectroscopic datasets.  None of the fields were chosen because of the presence of
known absorption systems, although 
an \ion{O}{6}-bearing Lyman limit system at $z=0.167$ toward
PKS0405--123 had been analyzed by our group previously \citep{cp00}.  
As such, the galaxies discovered by our
survey provide an unbiased sample that can then be searched for associated
hydrogen and metal-line absorption.

This LCO/WFCCD galaxy survey provides the galaxy sample for the
following analysis;  we consider the results from complementary surveys 
in the discussion that follows ($\S$~\ref{sec:discussion}).
Altogether, the LCO/WFCCD galaxy survey comprises 1198 galaxies with 
$0.005 < z < (z_{\rm em} - 0.01)$ with a median redshift of 0.18.  The
distribution of their luminosities peaks near $\approx 0.3 L^*$ and
extends to $0.01~L^*$ and $\approx 5L^*$.

\begin{deluxetable*}{crrcccccccc}
\tablewidth{0pc}
\tablecaption{New Targeted Search for \lya\ and \ovi\ Absorption Associated with Galaxies\label{tab:newabs}}
\tabletypesize{\footnotesize}
\tablehead{\colhead{Quasar} & \colhead{RA} & \colhead{DEC} & \colhead{$z_{\rm gal}$} 
& \colhead{$W^{\mlya}$} & \colhead{\nhi} 
&\colhead{$W^{\rm OVI}$} & \colhead{$N_{\rm OVI}$} & \colhead{UV Spectra}
\\
 & (J2000) & (J2000) & &  (\AA)  & & (m\AA)} 
\startdata
Q0026+1259&00:29:13.8 & +13:16:04       & 0.03290 & \dots & \dots &$<120$&$<14.0$&FUSE\\
&&&0.03930 & \dots & \dots &$<120$&$<14.0$&FUSE\\
&&&0.09610 & \dots & \dots &$<100$&$<14.0$&FUSE\\
&&&0.11250 & \dots & \dots &$<100$&$<14.0$&FUSE\\
PKS0405-123&04:07:48.4 & -12:11:37       & 0.04320 & $<0.03$&$<12.7$&\dots & \dots &STIS/E140M\\
&&&0.07910 & $<0.03$&$<12.7$&$< 40$&$<13.5$&FUSE, STIS/E140M\\
&&&0.08000 & $<0.03$&$<12.7$&$< 40$&$<13.5$&FUSE, STIS/E140M\\
&&&0.15220 & \dots & \dots &$< 50$&$<13.6$&STIS/E140M\\
&&&0.20300 & $<0.05$&$<13.0$&$< 30$&$<13.4$&STIS/E140M\\
&&&0.20890 & $<0.03$&$<12.7$&\dots & \dots &STIS/E140M\\
&&&0.24840 & $<0.05$&$<12.9$&$< 30$&$<13.4$&STIS/E140M\\
&&&0.28000 & $<0.03$&$<12.8$&\dots & \dots &STIS/E140M\\
&&&0.29510 & $<0.09$&$<13.2$&$< 30$&$<13.4$&STIS/E140M\\
&&&0.29760 & \dots & \dots &$< 50$&$<13.6$&STIS/E140M\\
&&&0.30990 & $<0.04$&$<12.9$&$< 40$&$<13.5$&STIS/E140M\\
&&&0.32030 & $<0.05$&$<13.0$&$< 30$&$<13.4$&STIS/E140M\\
&&&0.32520 & \dots & \dots &$< 50$&$<13.6$&STIS/E140M\\
&&&0.34150 & $<0.15$&$<13.5$&\dots & \dots &STIS/E140M\\
&&&0.36810 & $<0.05$&$<13.0$&\dots & \dots &STIS/E140M\\
&&&0.37850 & $<0.06$&$<13.0$&\dots & \dots &STIS/E140M\\
PG1004+130&10:07:26.1 & +12:48:56       & 0.00920 & $0.26\pm 0.03$& $13.9 \pm 0.15$& $< 50$&$<13.6$&FUSE, STIS/G140M\\
&&&0.03008 & \dots & \dots &$<200$&$<14.5$&FUSE\\
&&&0.07040 & \dots & \dots &$<100$&$<13.9$&FUSE\\
&&&0.16740 & \dots & \dots &$< 80$&$<13.8$&STIS/G140M\\
&&&0.19310 & \dots & \dots &$< 80$&$<13.8$&STIS/G140M\\
PG1216+069&12:19:20.9 & +06:38:38       & 0.00630 & $2.32\pm 0.06$& $19.3 \pm 0.05$& $<200$&$<14.2$&FUSE, STIS/E140M\\
&&&0.00810 & $<0.06$&$<13.1$&$<200$&$<14.2$&FUSE, STIS/E140M\\
&&&0.01262 & $0.35\pm 0.03$& $>14.1$&$<200$&$<14.2$&FUSE, STIS/E140M\\
&&&0.08150 & \dots & \dots &$<120$&$<14.0$&FUSE\\
&&&0.11840 & $<0.03$&$<12.7$&$<130$&$<14.1$&FUSE, STIS/E140M\\
&&&0.11910 & $<0.03$&$<12.7$&$<130$&$<14.1$&FUSE, STIS/E140M\\
&&&0.13540 & \dots & \dots &$<130$&$<14.1$&FUSE\\
&&&0.18100 & \dots & \dots &$< 80$&$<13.8$&STIS/E140M\\
&&&0.19170 & $<0.05$&$<12.9$&$< 40$&$<13.5$&STIS/E140M\\
&&&0.24640 & $<0.05$&$<12.9$&$< 40$&$<13.5$&STIS/E140M\\
&&&0.27950 & $<0.05$&$<12.9$&$< 60$&$<13.6$&STIS/E140M\\
&&&0.28000 & $<0.05$&$<12.9$&$< 60$&$<13.6$&STIS/E140M\\
PKS1302-102&13:05:33.0 & -10:33:19       & 0.02520 & $<0.10$&$<13.3$&$< 40$&$<13.5$&FUSE, STIS/E140M\\
&&&0.03650 & $<0.25$&$<15.0$&\dots & \dots &STIS/E140M\\
&&&0.05690 & $<0.35$&$<15.0$&$< 40$&$<13.5$&FUSE, STIS/E140M\\
&&&0.07120 & $<0.50$&$<16.0$&$< 40$&$<13.5$&FUSE, STIS/E140M\\
&&&0.07180 & $<0.50$&$<16.0$&$< 40$&$<13.5$&FUSE, STIS/E140M\\
&&&0.10620 & $<0.05$&$<12.9$&\dots & \dots &STIS/E140M\\
&&&0.13930 & $<0.04$&$<12.8$&\dots & \dots &STIS/E140M\\
&&&0.14290 & $<0.04$&$<12.9$&$< 40$&$<13.5$&FUSE, STIS/E140M\\
MRK1383&14:29:06.4 & +01:17:06.0     & 0.02990 & \dots & \dots &$< 40$&$<13.5$&FUSE\\
\enddata
\tablecomments{The \nhi\ value for the $z=0.0063$ absorber toward PG1216+069 is taken from \cite{tripp06}}. 
\end{deluxetable*}

\subsection{IGM Surveys for \lya\ and \ovi\ Absorption}
\label{sec:absline}

In the following IGM/galaxy analysis, we leverage published surveys for
\lya\ and \ovi\ absorption in the quasar spectra of our 
fields.  This corresponds to \npub\ sightlines with UV spectra
from the {\it HST}/GHRS, {\it HST}/STIS, and/or {\it FUSE} instruments
(Table~\ref{tab:fields2}).  We restrict our
analysis to surveys based on high spectral resolution data which generally provide
equivalent width sensitivity $\sigma(W)$ to a few tens of m\AA.   
Although the published surveys
occasionally disagree on the identification of a particular line
or on the confidence of its detection, we have not attempted to
reconcile these arguments.  Instead, we adopt all published detections
as bona-fide systems. When multiple  
groups have reported measurements for the same system, we have adopted the
values from publications in this order: 
[\lya] -- \cite{tripp08,ds08,cm09,dsr+06,pss04}; 
[\ovi] -- \cite{tripp08,tc08a,ds08,dsr+06}.
Because we rely primarily on published surveys (see below), 
the IGM results were derived independently of our galaxy survey.

For galaxies within several hundred kpc (up to 1\,Mpc for \lya) of
these sightlines, we have also done a focused search for associated \lya\
and \ovi\ absorption at their redshifts.
Specifically,
if there were no published measurement near the redshift of the
galaxy yet spectra exist we reanalyzed the data ourselves.
Furthermore, 
sufficient quality UV spectra exist for 
\lya\ and/or \ovi\ analysis 
in one of the LCO/WFCCD survey fields
yet none has been published (Q0026+1259).
We have also done a focused analysis in this sightline according to
galaxies discovered in the LCO/WFCCD survey.
To perform this new IGM
analysis, we used spectra reduced by the instrument pipelines
as described in \cite{cpc+08}.  We normalized the quasar continuum
with automated algorithms and measured equivalent widths (and errors)
with simple boxcar summation.  
If no absorption was present within $\pm 400 \, \mkms$ of the galaxy
redshift, we report a $3\sigma$ upper limit to the equivalent width
and column density (assuming the linear curve-of-growth approximation).
For positive detections,
we measured column densities
with the apparent optical depth method \citep{savage91}.  
In a few cases, the predicted wavelength of a \lya\ or \ovi\ line coincides
with Galactic ISM absorption or a coincident transition from another
absorption system.  In these events, we adopt the equivalent width of
the coincident line as an upper limit to the equivalent width.

The measurements for our IGM analysis are summarized in Table~\ref{tab:newabs}. 
Altogether, we have useful constraints on
\lya\ and/or \ovi\ absorption for \nanly\ of our WFCCD survey fields.

\section{Results}
\label{sec:results}


We take two complementary approaches to the analysis: (i) we examine
all galaxies that lie close to the sightlines for associated \lya\ and
\ovi\ absorption; and
(ii) we search the fields for galaxies associated with detected \lya\ and \ovi\
absorbers.  Each of these techniques has specific strengths and
weaknesses for interpreting the IGM/galaxy connection, as described
below. Together, however, they provide powerful insight into
the nature of the IGM and its interplay with galaxies at $z \sim 0$.

In the following, we establish the velocity criterion for associating a
galaxy with IGM absorption (or vice-versa) to be a velocity difference
$|\delta v| < 400 \, \mkms$.  This value was chosen to be large enough to account for error
in the galaxy redshifts \citep{ovi_paper4} and to allow for velocities
between the galaxy and gas that are characteristic of galactic
dynamics.  We emphasize that over $80\%$ of the
purported galaxy/absorber associations have $|\delta v| \le 250\,\mkms$ and that our
results would not change qualitatively if we adopted a 
value lower than 400\,\kms.

We examine the results with particular emphasis on the galaxy
luminosity. These were derived from the apparent
$R$-band magnitude, redshift, and spectral type
\citep[see][]{ovi_paper4}. 
Generally, we consider three luminosity intervals:
(1) $L<0.1L^*$, termed dwarf galaxies;
(2) $0.1 L^* < L < L^*$, termed sub-$L^*$ galaxies and;
(3) $L>L^*$, termed $L^*$ galaxies.  
For these definitions, 
we adopt the $r$-band value for $L^*$ 
as measured from the Sloan Digital Sky Survey which corresponds to $z
\approx 0.1$: $r^* = -20.44 + 5\log h_{100}$ \citep{blanton03}.
With the cosmology we have adopted, this yields $r^* = -21.12$\,mag.  
We are motivated to this approach by the expectation that galaxies
of a similar luminosity will, on average, trace similar environments
and have dark matter halos with common characteristics.  
Ideally, we might separate the galaxies by their dark matter halo
mass, as inferred from the galaxy's kinematics.  Unfortunately, our
spectra are of insufficient quality to discriminate on kinematics nor
do we have the necessary photometry to estimate stellar masses.
Regarding the galaxy spectral-type, we do make the distinction between
early-type (absorption-line dominated) and late-type (emission-line
dominated) galaxies based on our spectra \citep[see][for more
details]{ovi_paper4}.

When discussing the results, we
make a crude distinction between the virialized
halo of a galaxy defined by a virial radius \rvir\ and the
circumgalactic medium (CGM) that
surrounds it to a larger radius \rcgm.  The
motivation for this distinction is that galactic-scale processes (e.g.\
accretion, shock-heating, outflows) may be especially active within
the virialized halo, and therefore produce a medium that is qualitatively
different from the surrounding CGM.  
Furthermore gas at $r < \mrvir$ may have a much greater probability of
being gravitationally bound to the galaxy.  
At the same time, we doubt that
\rvir\ marks a sharp boundary of any sort.
Lastly, we note that a sightline which penetrates a galactic halo may
show absorption from the gas within it and the gas that surrounds it.

In standard numerical and analytic analysis, \rvir\ is predicted to
scale with the dark matter halo mass as $\mrvir \propto
M_h^{1/3}$ and with the circular velocity as $\mrvir \propto
v_c$.  If we were to adopt the observed $R$-band Tully-Fisher (TF) relation
$L \propto v_c^{3.4}$ \citep{courteau+07}, we would infer that \rvir\
scales as $\mrvir \propto L^{0.3}$.  More modern treatments compare
the clustering of galaxies as a function of luminosity to predictions
for dark matter halos and thereby derive a relation between $L$ and
$M_h$ \citep[e.g.][]{zcz07,zzw+10}.
These results, which are dominated by $L>L^*$ galaxies, imply $M_h
\propto L^{3.8}$ (in $R$-band) at high luminosity and a much flatter relation at low
luminosity (consistent with the higher $M/L$ ratio inferred for
low-mass galaxies).  Therefore, while most treatments --- empirical and
theoretical --- tend to agree that $\mrvir \approx 250$--$300$\,kpc for an
\lstar\ galaxy, various prescriptions yield very different estimates
for the virial radius of low luminosity galaxies.  In the following,
we adopt 
characteristic values of $\mrvir \approx 100$\,kpc for the dwarf
galaxies, $\mrvir \approx 160$\,kpc for a \subls\ galaxy, and 
$\mrvir \approx 250$\,kpc for an \lstar\ galaxy.  
This corresponds to a scaling relation:

\begin{equation}
\mrvir = r^*_{\rm vir} \ltp \frac{L}{L^*} \rtp^\beta
\label{eqn:rvir_pow}
\end{equation}
with $r^*_{\rm vir}= 250$\,kpc and $\beta = 0.2$.
These assignments and this relation should be considered crude
estimates and primarily serve to guide the discussion.



\subsection{Gas Associated with Galaxies at Low Impact Parameter}
\label{sec:step1}

In this subsection, we address the following question:
granted that a galaxy with a certain luminosity lies at a given impact parameter from
a quasar sightline, what is the probability of detecting \lya\ or
\ovi\ absorption at that same redshift (within a given velocity
interval $\delta v$) to a given equivalent width limit?  More
generally, we consider the distribution of equivalent widths and ionic
column densities as a function of the galaxy luminosity $L$ and impact
parameter $\rho$.  In principle,
such measurements reveal the physical conditions of gas in the
virialized halos of galaxies and in the circumgalactic
medium (CGM) that surrounds
them (e.g.\ filaments, the intragroup medium).
In turn, the measurements may inform whether the gas is physically
connected to the galaxy (e.g.\ gravitationally bound to its dark matter halo) 
and/or which galaxies may `produce' IGM absorption at specific equivalent widths and
column density.  

\begin{figure*}
\epsscale{0.6}
\includegraphics[height=6.5in,angle=90]{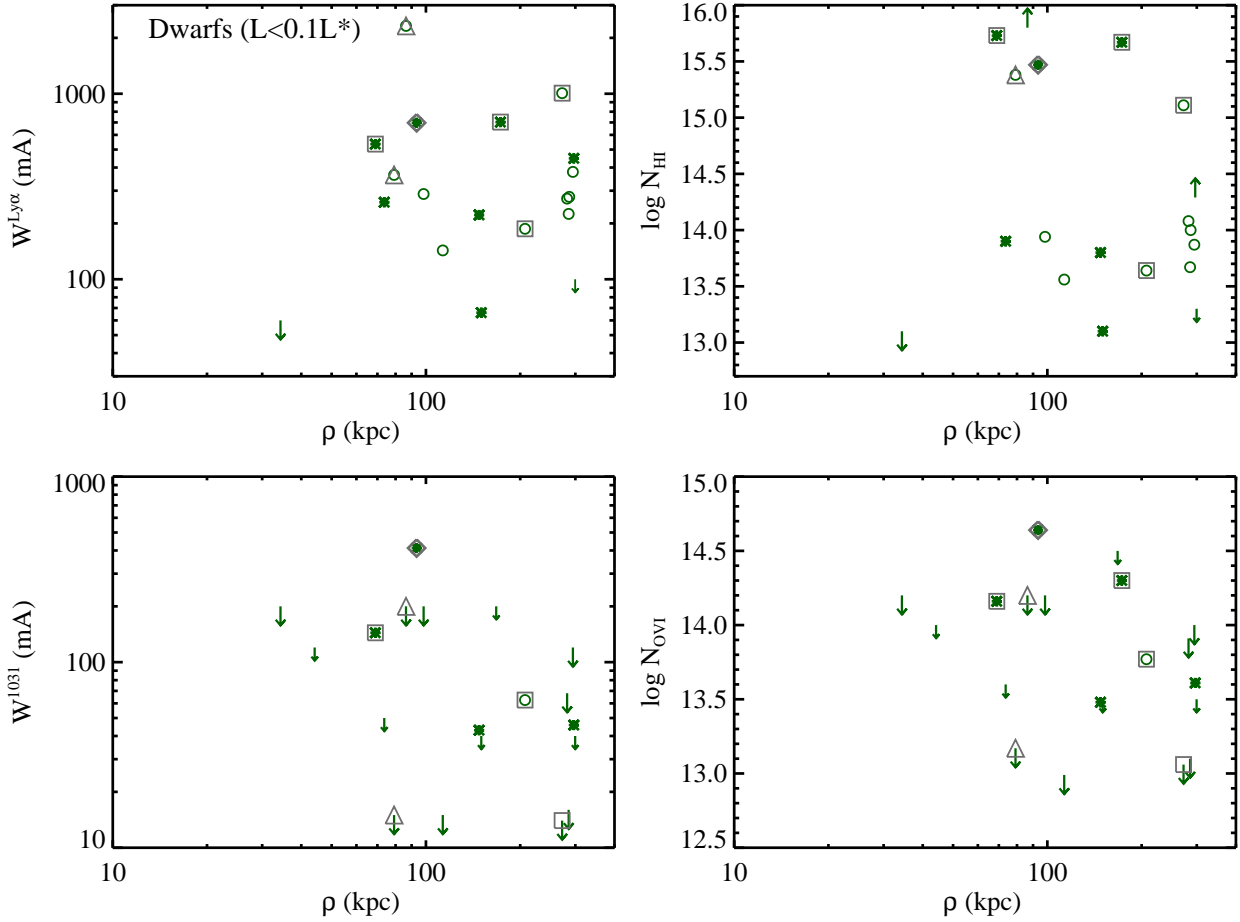}
\caption{The four panels show the measured equivalent widths and
  column densities of \lya\ and \ovi\ absorption for all of the dwarf
  galaxies ($L<0.1L^*$) in our LCO/WFCCD survey that lie within 300\,kpc of a quasar sightline (where high
  quality UV spectroscopy exists).   In each case, we report the
  maximum value for all \lya\ (or \ovi) absorption lines detected within 400\kms\ of the
  galaxy redshift.  Asterisks denote galaxies with
  late-type spectra and open circles signs denote early-type spectra
  (small/large arrows indicated limits for late/early-type galaxies).
  In cases where there is more than
  one dwarf galaxy within 400\,\kms\ of one another, we plot only
  the one closest to the sightline.  Upper limits denote non-detections or, in a few
  cases, unresolved blends with coincident absorption lines.  The gray
  boxes (diamonds) denote systems where an additional \subls\ (\lstar)
  galaxy lies within 150\,kpc of the sightline.  Remarkably, these
  include nearly all of the systems associated with strong \lya\
  absorption and positive \ovi\ detections.  The gray triangles,
  meanwhile, denote galaxies associated with Virgo.  Note the
  preponderance of \lya\ associations but the rare detection of \ovi\
  gas, especially if one ignores the dwarf galaxies with a brighter
  neighbor near the sightline.
}
\label{fig:dwarfs}
\end{figure*}

There is (at least) one significant challenge to this analysis:
galaxies cluster with one another.  Therefore, if a galaxy lies at 
close impact parameter to a sightline there is a significant
probability that one or more additional galaxies will also lie close
to the same sightline.  
If we then detect IGM absorption at that redshift, there is an ambiguous
association between the gas and each galaxy.
One may be able to design
an experiment that 
studied only isolated galaxies, but this would likely be
unrepresentative of the full galaxy population.  This issue is further
complicated by the imperfections of our LCO/WFCCD survey. These
include 
(i) the magnitude limit which implies sensitivity to fainter galaxies only
at lower redshifts;
(ii) the limited field-of-view which precludes a search for galaxies with
$\rho \lesssim 100$\,kpc at the lowest redshifts; and 
(iii) the varying
sensitivity of the UV spectrometers with wavelength which dictates the
equivalent width limit to \lya\ and \ovi\ absorption as a function of
redshift.   In the following, we emphasize these limitations where appropriate.

To (partially) address the ambiguity of multiple galaxies at a given
redshift,\footnote{Previous authors have introduced criteria based on
  galaxy clustering to select the galaxy of interest when two or more
  with similar redshift are located near the sightline
  \citep{chenetal01}.  In general, this amounts to selecting the galaxy
  with smallest impact parameter \citep{mj06}.}
we perform the analysis separately for dwarf, \subls, and
\lstar\ galaxies.
Initially, we examine all galaxies at
impact parameter $\rho < 300$\,kpc from the sightlines.
This distance was arbitrarily chosen to exceed the virial radius of
the galaxies yet to be small enough to primarily focus the analysis on
individual galaxies. 
When two or more galaxies are separated by $|\delta v| < 400 \mkms$
and have similar luminosity, we
only present the one closest to the sightline (this occurs very rarely).   

\begin{deluxetable*}{crrcccccccccccccc}
\tablewidth{0pc}
\tablecaption{Dwarf Galaxies within 300 kpc of a QSO Sightline\label{tab:dwarfs}}
\tabletypesize{\scriptsize}
\tablehead{\colhead{Field} & \colhead{RA} & \colhead{DEC} & \colhead{$z_{\rm gal}$} & \colhead{T$^a$} & \colhead{$\rho$} & 
\colhead{$L$} & \colhead{$z_{\rm abs}^{\mlya}$} & \colhead{$W^{\mlya}$} &
\colhead{\nhi} & \colhead{Rf} & \colhead{$z_{\rm abs}^{\rm OVI}$} &
\colhead{$W^{\rm OVI}$} & \colhead{$N_{\rm OVI}$} & \colhead{Rf} &
\colhead{$\rho_{\rm n}^b$} & \colhead{$L_{\rm n}^b$} \\
 & (J2000) & (J2000) & & & (kpc) & ($L^{*}$) & & (m\AA) & & & & (m\AA) &&&&($L^{*}$) } 
\startdata
Q0026+1259 & 00:29:09.3 & +13:16:29 & 0.0329 & L &   43 &  0.018 & \dots & \dots & \dots & &0.0329 & $< 120$&$<14.0$& 9 & \dots & \dots \\
TONS180 & 00:57:04.0 & --22:26:51 & 0.0234 & L&  147 &  0.019 & 0.0234 &$ 222$&$13.80$& 5 & 0.0234 & $  43$&$13.48$& 4 & \dots & \dots \\
PKS0405-123 & 04:07:48.3 & --12:11:02 & 0.1670 & L&   93 &  0.080 & 0.1671 &$ 697$&$15.47$& 3 & 0.1669 & $ 412$&$14.64$& 1 & 115 & 2.14\\
PG1004+130 & 10:07:06.5 & +12:53:52 & 0.0092 & L&   73 &  0.007 & 0.0092 &$ 260$&$13.90$& 9 & 0.0092 & $<  50$&$<13.6$& 9 & \dots & \dots \\
  & 10:07:30.8 & +12:53:50 & 0.0297 & L&  167 &  0.051 & \dots & \dots & \dots & &0.0301 & $< 200$&$<14.50$& 9 & \dots & \dots \\
HE1029-140 & 10:31:41.4 & --14:12:49 & 0.0508 & E &  286 &  0.033 & 0.0516 &$ 278$&$14.00$& 5 & \dots & \dots & \dots & & \dots & \dots \\
PG1116+215 & 11:19:03.1 & +21:16:24 & 0.0594 & E&  207 &  0.024 & 0.0593 &$ 187$&$13.64$& 1 & 0.0593 & $  63$&$13.77$& 1 & 124 & 0.10\\
PG1211+143 & 12:14:06.9 & +14:04:38 & 0.0520 & L&  172 &  0.089 & 0.0510 &$ 703$&$15.67$& 3 & 0.0513 & \ldots&$14.30$& 4 & 131 & 0.83\\
  & 12:14:13.9 & +14:03:31 & 0.0646 & L&   68 &  0.084 & 0.0645 &$ 535$&$15.73$& 3 & 0.0645 & $ 144$&$14.16$& 3 & 141 & 0.74\\
PG1216+069 & 12:18:38.7 & +06:42:29 & 0.0067 & E&   86 &  0.029 & 0.0063 &$2320$&$19.30$& 9 & 0.0063 & $< 200$&$<14.2$& 9 & \dots & \dots \\
  & 12:19:20.7 & +06:42:19 & 0.0081 & E&   34 &  0.008 & 0.0081 &$<  60$&$<13.1$& 9 & 0.0081 & $< 200$&$<14.2$& 9 & \dots & \dots \\
  & 12:19:03.8 & +06:33:43 & 0.0132 & E&   98 &  0.014 & 0.0126 &$ 288$&$13.94$& 6 & 0.0126 & $< 200$&$<14.2$& 9 & \dots & \dots \\
  & 12:19:14.5 & +06:35:33 & 0.0805 & E&  294 &  0.061 & 0.0805 &$ 379$&$13.87$& 3 & 0.0815 & $< 120$&$<14.0$& 9 & \dots & \dots \\
3C273 & 12:29:50.6 & +02:01:54 & 0.0062 & E&   79 &  0.006 & 0.0053 &$ 365$&$15.38$& 3 & 0.0053 & $<  15$&$<13.2$& 4 & \dots & \dots \\
PKS1302-102 & 13:05:58.0 & --10:24:50 & 0.0252 & L&  299 &  0.009 & 0.0252 &$< 100$&$<13.3$& 9 & 0.0252 & $<  40$&$<13.5$& 9 & \dots & \dots \\
MRK1383 & 14:29:42.4 & +01:17:51 & 0.0281 & E&  285 &  0.009 & 0.0282 &$ 225$&$13.67$& 3 & 0.0283 & $<  16$&$<13.1$& 4 & \dots & \dots \\
  & 14:28:58.4 & +01:13:06 & 0.0299 & L&  150 &  0.023 & 0.0299 &$  66$&$13.10$& 5 & 0.0299 & $<  40$&$<13.5$& 9 & \dots & \dots \\
FJ2155-0922 & 21:54:56.8 & --09:27:24 & 0.0504 & E&  282 &  0.045 & 0.0515 &$ 272$&$14.08$& 3 & 0.0501 & $<  68$&$<13.9$& 4 & \dots & \dots \\
  & 21:54:47.4 & --09:23:05 & 0.0779 & L&  296 &  0.053 & 0.0777 &$ 448$&$>14.3$& 1 & 0.0777 & $  46$&$13.61$& 1 & \dots & \dots \\
  & 21:54:52.3 & --09:24:38 & 0.0808 & E&  272 &  0.070 & 0.0808 &$1007$&$15.11$& 3 & 0.0807 & $<  14$&$<13.1$& 4 &  33 & 0.87\\
PKS2155-304 & 21:58:30.5 & --30:11:02 & 0.0169 & E&  113 &  0.002 & 0.0170 &$ 143$&$13.56$& 3 & 0.0167 & $<  15$&$<13.0$& 4 & \dots & \dots \\
\enddata
\tablecomments{Upper limits correspond to $2\sigma$ constraints.}
\tablenotetext{a}{Spectral type (E=Early-type; L=Late-type; see text for the quantitative definition).}
\tablenotetext{b}{Impact parameter and luminosity of the brightest galaxy ($L>0.1L^*$) within 200 kpc and $|\delta v| < 400 \mkms$ (if any).}
\tablerefs{1: \cite{tripp08}; 2: \cite{tc08a}; 3: \cite{ds08}; 4: \cite{dsr+06}; 5: \cite{pss04}; 6: \cite{cm09}; 9: This paper.}
\end{deluxetable*}

\noindent
{\bf Dwarf Galaxies:}
Consider first the $L<0.1L^*$ galaxies (a.k.a.\ the dwarf galaxies); 
Table~\ref{tab:dwarfs} lists the sample of systems at low impact
parameter from our WFCCD survey. 
The table also lists the strongest \lya\ and \ovi\ absorbers within
$|\delta v| \le 400\,\mkms$ of each
galaxy, or an upper limit to the equivalent width and column density
for non-detections.  We also tabulate the closest galaxy with
$L > 0.1L^*$ if one (or more) lies within 300\,\kms\ of the dwarf galaxy's
redshift and 200\,kpc of the sightline. 

Figure~\ref{fig:dwarfs} summarizes the absorption characteristics of
the sample as a function of impact parameter.  Regarding \ion{H}{1}
absorption, we note the positive detection of \lya\ absorption for
16/19 ($84\%$) of the galaxies to $\mwlya = 100$\,m\AA\ or $\mnhi =
10^{13.5} \cm{-2}$.   
We emphasize that unlike the high-$z$ 
universe where the \lya\ forest is sufficiently `dense' that one
identifies a line every $\sim 100\mkms$, the probability of randomly
associating a galaxy with an absorber is small at $z \sim 0$.
Taking PG1116+215 as a representative sightline, the \lya\
lines reported by \cite{ds08} cover $\approx 33\%$ of the
sightline for $|\delta v| < 400 \, \mkms$ to $\mnhi \approx 10^{13}
\cm{-2}$.  Restricting to \lya\ lines with $\mnhi \ge 10^{13.5}
\cm{-2}$ (Figure~\ref{fig:dwarfs}), which is more characteristic of the gas associated with the
dwarf galaxies, we find only $\approx 10\%$ of the sightline is
covered. As such, if we randomly assigned redshifts
to the galaxies with values uniformly sampling $z = [0, z_{\rm em}]$,
the probability of recovering 16/19 associations with \lya\ lines
having $\mnhi > 10^{13.5} \cm{-2}$ is negligible.
We conclude, therefore, that \lya\ absorption traces dwarf galaxies to
impact parameters of 300\,kpc or greater with a nearly unit covering
fraction to $\mwlya = 100$\,m\AA\ ($\mnhi = 10^{13.5} \cm{-2}$).  

Inspecting Figure~\ref{fig:dwarfs} we note that the majority of
dwarf galaxy-\lya\ associations are at an impact parameter
that is well beyond the presumed virial radius of $\mrvir \approx 100$\,kpc.
We infer, therefore, that the connection between these galaxies and \ion{H}{1}
absorption is not driven (generally) by galactic-scale phenomena, e.g.\ tidal
debris, accreting gas, outflows.
\ion{H}{1} gas also arises in the extended CGM of these galaxies.
The direct implication is that the
two phenomena (\lya\ absorption and dwarf galaxies) are simply
highly correlated tracers of the same large-scale overdensity in the universe
(e.g.\ a filament).  
This assertion is further supported by the
fact that there is no strong correlation between the impact parameter
and the \lya\ absorption (in column density or equivalent width).

It is also evident from Figure~\ref{fig:dwarfs} that a subset of the
dwarf galaxies are associated with relatively strong \lya\ absorbers,
$W^{\rm Ly\alpha} > 0.5$\AA\ and $\mnhi > 10^{15} \cm{-2}$.  However,
most of these strong absorbers may also be associated to a brighter 
galaxy ($L>0.1L^*$) within 200\,kpc of the sightline.  We have
overplotted on the points gray boxes (diamonds) to indicate
dwarf galaxies where an additional sub-$L^*$ (\lstar) galaxy lies within
150\,kpc of the sightline.  With the exception of the two galaxies
associated with Virgo (at $z\approx 0.006$ toward PG1216+069 and
3C273; marked with gray triangles), all of the dwarf galaxies with
associated strong \lya\
absorption also have a brighter galaxy near the sightline.  
Furthermore, there are no cases where a brighter galaxy is nearby and
the \lya\ equivalent width is low.
Although limited by small number statistics, these results suggest that
the strongest \ion{H}{1} absorbers are preferentially associated with
a $L> 0.1 L^*$ galaxy instead of the dwarf.

Lastly, with regards to dwarf galaxies and \ion{H}{1} absorption,
we consider the two non-detections in \lya.  One case is located at
nearly 300\,kpc from PKS1302--102 and we may dismiss it because of the
large impact parameter.  The other example, however, is for a
$z=0.0081$ early-type galaxy that has the lowest impact parameter ($\rho = 34$\,kpc)
of the entire sample!  
Furthermore, it is the only example we have where the sightline is
certain to intersect the galaxy's virialized halo.
We identify no other galaxies at this redshift
and only note that it might be associated with the extreme outskirts of
Virgo.  Either way, this non-detection stresses that
the IGM/galaxy connection shows special cases that
contradict dominant trends.

The lower panels of Figure~\ref{fig:dwarfs} present the equivalent
width of \ion{O}{6}~1031 (\wovi; when UV spectral coverage exists) and an
estimate of the O$^{+5}$ column density \novi\ for the strongest absorber
within 300\,\kms\ of each galaxy.   
In contrast to \lya, the results are dominated by non-detections.
Furthermore, 3 of 5 systems with a positive detection (3/4
for those with $\mnovi > 10^{14} \cm{-2}$) also
show a brighter galaxy ($L>0.1L^*$) within 150\,kpc of the sightline.
Ignoring these systems,
0/13 of `isolated' dwarf galaxies show a positive detection to an equivalent
width limit of 200\,m\AA\ and only 2/8 for a 50\,m\AA\ limit.
We conclude that the CGM of dwarf galaxies at impact parameters of 50--300\,kpc  
rarely shows \ovi\ absorption characteristic of
current UV surveys.  
The CGM of dwarf galaxies does give rise to nearly
ubiquitous \ion{H}{1} absorption, but has
insufficient surface density,
metallicity, and/or the physical conditions (i.e.\ density, temperature)
needed to to also exhibit significant \ovi\ absorption.

\begin{figure*}
\includegraphics[height=6.5in,angle=90]{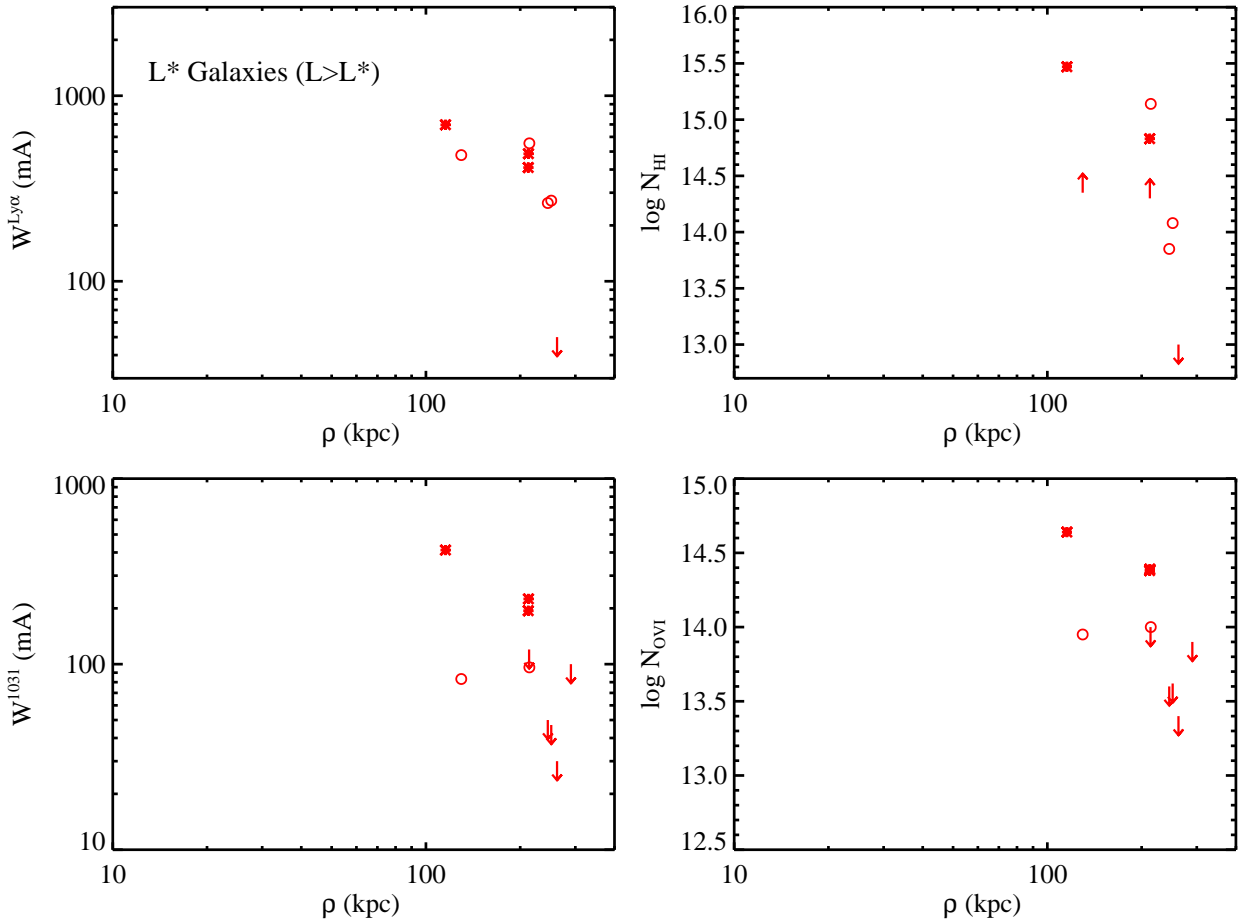}
\caption{Same as Figure~\ref{fig:dwarfs} but for \lstar\ galaxies.
  Note the high covering fraction to \lya\ absorption (with column
  densities typically exceeding $10^{14} \cm{-2}$) but the lack of
  \ovi\ detections for $\rho > 225$\,kpc to sensitive limits.  The
  results suggest that \ovi\ gas near \lstar\ galaxies is primarily
  associated with their virialized halos but not their extended CGM
  environments. 
}
\label{fig:lstar}
\end{figure*}

In summary, the circumgalactic medium of dwarf galaxies at $\rho
\gtrsim 100$\,kpc is characterized by a high incidence of moderate
strength
\lya\ absorption ($\mwlya \gtrsim 200$m\AA) and a low incidence of \ovi\
absorption to modest equivalent width limits. 
Our survey, unfortunately, offers few probes of dwarf galaxies on
smaller scales, i.e.\ with sightlines intersecting their dark matter
halos.  This assessment awaits a dedicated survey of such
galaxies.\footnote{An ongoing Cycle~18 HST/COS survey by J. Tumlinson
  (GO 12248)
  is designed to address this directly, although it will not include
  high S/N spectra of the \ovi\ doublet.}

\begin{deluxetable*}{crrccccccccccccc}
\tablewidth{0pc}
\tablecaption{$L>L^*$ Galaxies within 300 kpc of a QSO Sightline\label{tab:lstar}}
\tabletypesize{\scriptsize}
\tablehead{\colhead{Field} & \colhead{RA} & \colhead{DEC} & \colhead{$z_{\rm gal}$} & \colhead{T$^a$} & \colhead{$\rho$} & 
\colhead{$L$} & \colhead{$\mathcal{N}^b$} & \colhead{$z_{\rm abs}^{\mlya}$} & \colhead{$W^{\mlya}$} &
\colhead{\nhi} & \colhead{Rf.} & \colhead{$z_{\rm abs}^{\rm OVI}$} &
\colhead{$W^{\rm OVI}$} & \colhead{$N_{\rm OVI}$} & \colhead{Rf.} \\
 & (J2000) & (J2000) & & & (kpc) & ($L^{*}$) & & & (m\AA) & & & & (m\AA) }
\startdata
Q0026+1259 & 00:29:15.36 & +13:20:57.0 & 0.0393 & E &  213 &  1.384 &  0 & \dots & \dots & \dots & \dots &0.0393 & $< 120$&$<14.0$& 9\\
PKS0405-123 & 04:07:51.28 & --12:11:38.3 & 0.1670 & L  &  115 &  2.143 &  1 & 0.1671 &$ 697$&$15.47$& 3 & 0.1669 & $ 412$&$14.64$& 1\\
  & 04:07:42.79 & --12:11:33.1 & 0.2030 & E&  262 &  1.070 &  9 & 0.2030 &$<  50$&$<13.00$& 9 & 0.2030 & $<  30$&$<13.4$& 9\\
  & 04:07:50.69 & --12:12:25.2 & 0.2976 & E&  245 &  1.878 &  7 & 0.2977 &$ 264$&$13.85$& 3 & 0.2976 & $<  50$&$<13.6$& 9\\
  & 04:07:45.96 & --12:11:09.9 & 0.3612 & E&  213 &  6.604 &  5 & 0.3608 &$ 554$&$15.14$& 3 & 0.3616 & $  96$&$14.00$& 1\\
PG1004+130 & 10:07:34.55 & +12:52:09.5 & 0.0704 & E&  290 &  2.404 &  0 & \dots & \dots & \dots & \dots &0.0704 & $< 100$&$<13.9$& 9\\
PG1116+215 & 11:19:06.73 & +21:18:29.3 & 0.1383 & E&  129 &  2.039 &  4 & 0.1385 &$ 479$&$>14.35$& 1 & 0.1385 & $  83$&$13.95$& 1\\
PKS1302-102 & 13:05:20.22 & --10:36:30.4 & 0.0426 & L &  212 &  2.577 &  1 & 0.0422 &$ 410$&$14.83$& 3 & 0.0422 & $ 194$&$14.38$& 3\\
FJ2155-0922 & 21:54:56.64 & --09:18:07.9 & 0.0517 & E&  251 &  1.427 &  0 & 0.0515 &$ 272$&$14.08$& 3 & 0.0514 & $<  47$&$<13.6$& 4\\
  & 21:55:06.53 & --09:23:25.2 & 0.1326 & L &  212 &  1.626 &  2 & 0.1324 &$ 487$&$>14.30$& 1 & 0.1324 & $ 225$&$14.39$& 1\\
\enddata
\tablenotetext{a}{Spectral type (E=Early-type; L=Late-type; see text for the quantitative definition).}
\tablenotetext{b}{Number of additional galaxies within 3\,Mpc of the sightline,  400\,\kms\ of this galaxy, and having $L>0.1L^*$.}
\tablerefs{1: \cite{tripp08}; 2: \cite{tc08a}; 3: \cite{ds08}; 4: \cite{dsr+06}; 5: \cite{pss04}; 6: \cite{cm09}; 9: This paper.}
\end{deluxetable*}


\noindent
{\bf \lstar\ Galaxies:}
In Figure~\ref{fig:lstar}, we plot the same quantities versus impact
parameter but now for $L>L*$ galaxies (Table~\ref{tab:lstar}).
Our survey has only a modest sample of such bright galaxies
within 300\,kpc of the sightlines which reflects, of
course, their lower comoving number density. 
Similar to the dwarf galaxies, nearly every $L^*$
galaxy exhibits significant \lya\ absorption.\footnote{  
The only notable
exception is an $L \approx 1.0 L*$ galaxy with an early-type spectrum located
at an impact parameter of $\rho \approx 250$\,kpc from PKS0405--123
($z=0.203$).
Perhaps not coincidentally, it also is associated with one of the most significant
galaxy overdensities identified in our survey.}
This even includes 
luminous, early-type galaxies which are likely gas-poor on
galactic-scales. The average
equivalent width at $\rho = 100$--$200$\,kpc is $W^{\rm Ly\alpha} \approx
500$\,m\AA\ and $\approx 400$\,m\AA\ at $\rho = 200$--$300$\,kpc, 
twice that observed for
the dwarf galaxies (ignoring dwarfs with bright
neighbors; Table~\ref{tab:dwarfs}).  
Similarly, the \nhi\ values of the $L^*$ galaxies exceed
those for the dwarfs. 
This implies that the \ion{H}{1} gas near a typical \lstar\ galaxy has a
higher total surface density, neutral fraction, and/or a more extreme
velocity field.  Such trends with
luminosity have been revealed previously \citep[e.g.][]{chenetal01};
presumably, they are a simple consequence of these
galaxies having greater mass.

Regarding \ovi\ absorption, the $L^*$ galaxies show a greater fraction
(5/10) of positive detections than dwarf galaxies.  Furthermore, there is
a possible trend with impact
parameter: all of the detections occur within $\rho = 225$\,kpc, which
corresponds (roughly) to the virial radii of these luminous
galaxies. The figure
also reveals that every emission-line galaxy exhibits a positive
\ovi\ detection; all of the non-detections are associated with 
early-type spectra.  
A preference for \ovi\ absorption to occur in late-type galaxies was 
previously reported by \cite{cm09} for their modest sample
of galaxy/\ovi\ associations (Tumlinson et al., in prep.). 
Unfortunately, all of our
non-detections also occur at large $\rho$ and, therefore, we cannot 
separate the effects of impact parameter and spectral type with this
sample.  Nevertheless, the results demonstrate clearly that the
extended CGM of \lstar\ galaxies (at least those with early-type
spectra)
is frequently characterized by the non-detection of \ovi\ gas.

\begin{figure*}
\includegraphics[height=6.5in,angle=90]{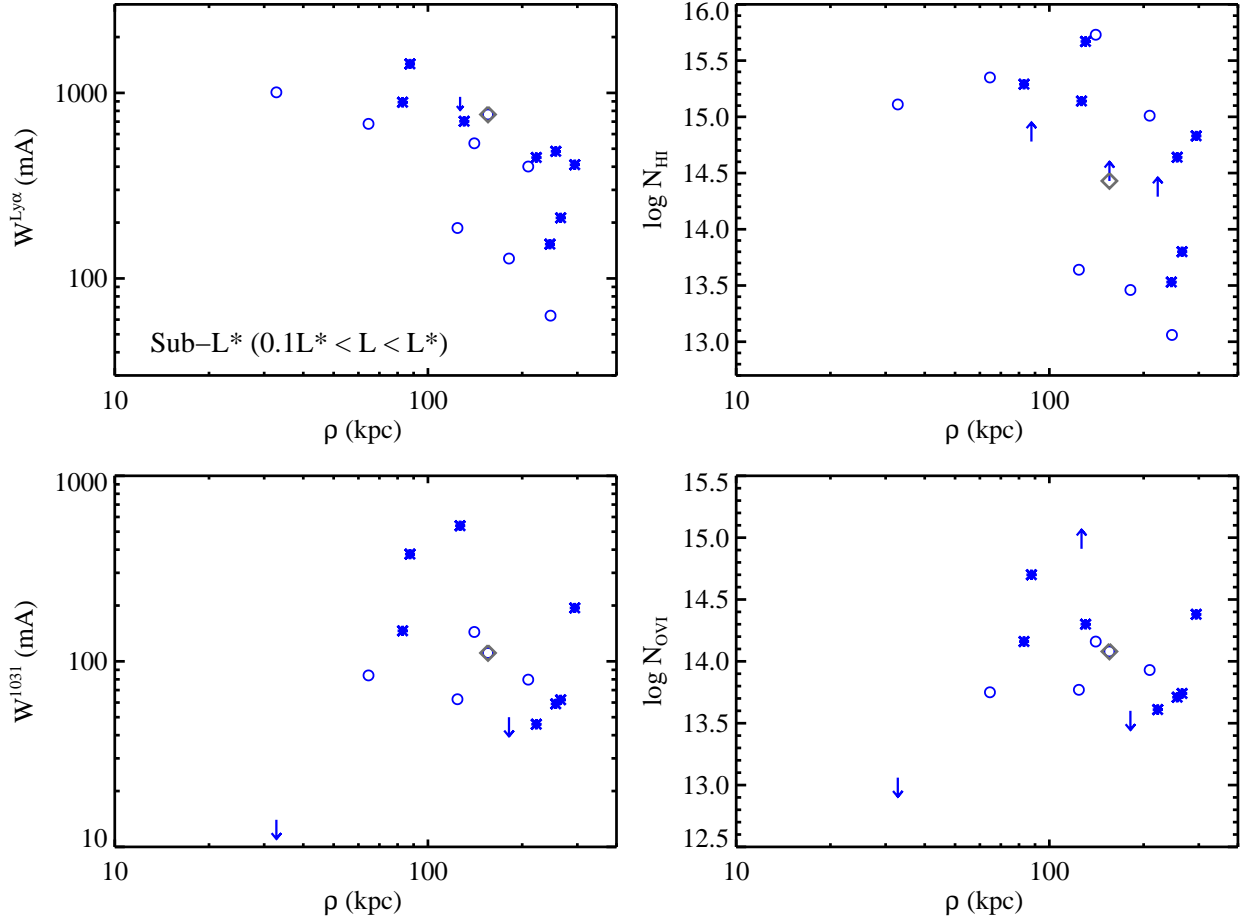}
\caption{Same as for Figure~\ref{fig:dwarfs} but for \subls galaxies.
One notes a strong correlation between \wlya\ and impact parameter for
these galaxies and also an apparent division at $\rho \approx
100$\,kpc between $\mwlya \gtrsim 1$\AA\ and lines exhibiting a large scatter of lower
values.  Intriguingly, this division occurs near the expected virial
radius for
these galaxies.  In contrast to the dwarf and \lstar\ galaxies
(Figures~\ref{fig:dwarfs}--\ref{fig:lstar}), we observe a very high
detection rate for \ovi\ at all impact parameters.  This indicates
the extended CGM of \subls\ galaxies has a high covering fraction to
modest \ovi\ absorption.
The gray diamond indicates the one \subls\ galaxy in our sample that
has an \lstar\ neighbor
within 150\,kpc of the
sightline.  
}
\label{fig:subls}
\end{figure*}

\clearpage

\begin{deluxetable*}{crrccccccccccccc}[ht]
\tablewidth{0pc}
\tablecaption{\subls\ Galaxies ($0.1L^*<L<L^*$) within 300 kpc of a QSO Sightline\label{tab:subls}}
\tabletypesize{\scriptsize}
\tablehead{\colhead{Field} & \colhead{RA} & \colhead{DEC} & \colhead{$z_{\rm gal}$} & \colhead{T$^a$} & \colhead{$\rho$} & 
\colhead{$L$} & \colhead{$\mathcal{N}^b$} & \colhead{$z_{\rm abs}^{\mlya}$} & \colhead{$W^{\mlya}$} &
\colhead{\nhi} & \colhead{Rf.} & \colhead{$z_{\rm abs}^{\rm OVI}$} &
\colhead{$W^{\rm OVI}$} & \colhead{$N_{\rm OVI}$} & \colhead{Rf} \\
 & (J2000) & (J2000) & & & (kpc) & ($L^{*}$) & & & (m\AA) & & & & (m\AA) }
\startdata
TONS180 & 00:57:08.52 & --22:18:29.6 & 0.0456 & L  &  265 &  0.365 &  1 & 0.0456 &$ 212$&$13.80$& 5 & 0.0456 & $  62$&$13.74$& 4\\
PKS0312-77 & 03:12:01.76 & --76:55:17.7 & 0.0594 & L &  245 &  0.292 &  0 & 0.0595 &$ 153$&$13.53$& 3 & \dots & \dots & \dots & \\
  & 03:11:57.89 & --76:51:55.6 & 0.2026 & L &  126 &  0.634 &  2 & 0.2027 &$< 952$&$15.14$& 3 & 0.2027 & $ 538$&$>14.9$& 1\\
PKS0405-123 & 04:07:58.12 & --12:12:24.7 & 0.0965 & L &  255 &  0.150 &  8 & 0.0966 &$ 484$&$14.64$& 3 & 0.0966 & $  59$&$13.71$& 3\\
  & 04:07:44.03 & --12:12:09.5 & 0.1532 & E &  181 &  0.980 &  1 & 0.1522 &$ 128$&$13.46$& 3 & 0.1522 & $<  50$&$<13.6$& 9\\
  & 04:07:48.40 & --12:12:10.8 & 0.3520 & U  &  156 &  0.273 &  2 & 0.3509 &$ 380$&$14.04$& 3 & 0.3509 & $  24$&$13.35$& 3\\
PG1116+215 & 11:19:05.55 & +21:17:33.3 & 0.0600 & E&  124 &  0.104 &  5 & 0.0593 &$ 187$&$13.64$& 1 & 0.0593 & $  63$&$13.77$& 1\\
  & 11:19:12.20 & +21:18:52.0 & 0.1660 & E&  155 &  0.336 & 15 & 0.1655 &$ 765$&$>14.4$& 1 & 0.1655 & $ 111$&$14.08$& 1\\
PG1211+143 & 12:14:09.54 & +14:04:21.3 & 0.0511 & L &  130 &  0.829 &  8 & 0.0510 &$ 703$&$15.67$& 3 & 0.0513 & \ldots&$14.30$& 4\\
  & 12:14:19.86 & +14:05:10.3 & 0.0644 & E&  140 &  0.740 &  0 & 0.0645 &$ 535$&$15.73$& 3 & 0.0645 & $ 144$&$14.16$& 3\\
PG1216+069 & 12:19:23.45 & +06:38:20.3 & 0.1241 & L &   87 &  0.650 &  6 & 0.1236 &$1433$&$>14.8$& 1 & 0.1236 & $ 378$&$14.70$& 1\\
PKS1302-102 & 13:05:25.65 & --10:39:23.5 & 0.0420 & L &  294 &  0.328 &  1 & 0.0422 &$ 410$&$14.83$& 3 & 0.0422 & $ 194$&$14.38$& 3\\
  & 13:05:32.19 & --10:33:56.9 & 0.0936 & E&   64 &  0.195 &  7 & 0.0949 &$ 681$&$15.35$& 3 & 0.0948 & $  84$&$13.75$& 4\\
  & 13:05:35.30 & --10:33:24.2 & 0.1453 & L &   83 &  0.396 &  2 & 0.1453 &$ 890$&$15.29$& 3 & 0.1453 & $ 146$&$14.16$& 4\\
  & 13:05:34.97 & --10:34:22.6 & 0.1917 & E&  209 &  0.888 &  2 & 0.1916 &$ 401$&$15.01$& 3 & 0.1916 & $  80$&$13.93$& 1\\
FJ2155-0922 & 21:54:50.87 & --09:22:33.3 & 0.0788 & L &  221 &  0.157 & 22 & 0.0777 &$ 448$&$>14.3$& 1 & 0.0777 & $  46$&$13.61$& 1\\
  & 21:54:59.96 & --09:22:24.7 & 0.0810 & E&   32 &  0.872 & 10 & 0.0808 &$1007$&$15.11$& 3 & 0.0807 & $<  14$&$<13.1$& 4\\
  & 21:55:01.62 & --09:20:47.0 & 0.1555 & E&  246 &  0.441 &  1 & 0.1549 &$  63$&$13.06$& 3 & \dots & \dots & \dots & \dots \\
\enddata
\tablenotetext{a}{Spectral type (E=Early-type, L=Late-type, U=unknown; see text for the quantitative definition).}
\tablenotetext{b}{Number of additional galaxies within 3\,Mpc of the sightline,  400\,\kms\ of this galaxy, and having $L>0.1L^*$.}
\tablerefs{1: \cite{tripp08}; 2: \cite{tc08a}; 3: \cite{ds08}; 4: \cite{dsr+06}; 5: \cite{pss04}; 6: \cite{cm09}; 9: This paper.}
\end{deluxetable*}

\noindent
{\bf Sub-$L^*$ Galaxies:}
Our final galaxy subset is the sub-$L^*$ galaxies
($0.1L^* < L < L^*$; Table~\ref{tab:subls}).
Figure~\ref{fig:subls} reveals a 100\%\ detection rate for \ion{H}{1}
gas to $\rho = 300$\,kpc for these galaxies.  
In contrast to the other galaxy subsets, the \lya\ equivalent widths
of the \subls\ galaxies show
a very obvious trend with impact parameter.  
A Spearman's test reports that the null hypothesis of no correlation
is ruled out at $> 99.5\%$\,c.l.
There is also the hint of a 
division in the \wlya\ values at $\rho \approx 100$\,kpc, i.e.\ 
between exclusively large equivalent widths ($\mwlya > 1$\,\AA\ for
$\rho < 100$\,kpc) and a
large scatter of primarily lower values for $\rho > 100$\,kpc (Table~\ref{tab:gastat}). 
As this impact parameter roughly coincides with the 
virial radius expected for \subls\ galaxies, if confirmed, 
it may be related to 
galactic-scale processes at $\rho < \mrvir$ whereas the \wlya\
values at larger $\rho$ trace the physical properties of the
surrounding (unvirialized) CGM.

\begin{deluxetable*}{cccccccccccccc}
\tablewidth{0pc}
\tablecaption{Galaxy/Absorber Statistics \label{tab:gastat}}
\tabletypesize{\scriptsize}
\tablehead{\colhead{Luminosity} & \colhead{$\rho$} &
  \colhead{$\mathcal{N}$} & \colhead{Med(\wlya)} & \colhead{RMS(\wlya)}
& \colhead{$\mathcal{N}$} & \colhead{M(\nhi)} & \colhead{R(\nhi)}
& \colhead{$\mathcal{N}$} & \colhead{M(\wovi)} & \colhead{R(\wovi)}
& \colhead{$\mathcal{N}$} & \colhead{M(\novi)} & \colhead{R(\novi)} \\
 & (kpc) & & (\AA) & (\AA) & &&&& (m\AA) & (m\AA) }
\startdata
$L < 0.1L^*$ \\
& 0--100 & 7&0.37&0.77&7&15.4& 2.0&8&$<$200&121&8&$<$14.2& 0.4\\
& 100--200 & 4&0.22&0.29&4&13.8& 1.1&4&$<$ 43& 84&5&$<$13.5& 0.6\\
& 200--300 & 8&0.28&0.28&8&14.0& 0.5&7&$<$ 45& 36&7&$<$13.6& 0.4\\
$0.1 < L < L^*$ \\
& 0--100 & 4&1.01&0.32&4&15.3& 0.3&4&146&158&4&14.2& 0.7\\
& 100--200 & 7&0.54&0.31&7&14.4& 0.9&6&111&192&7&14.1& 0.5\\
& 200--300 & 7&0.40&0.16&7&14.3& 0.7&5& 62& 60&5&13.7& 0.3\\
$L > L^*$ \\
& 100--200 & 2&0.70&0.15&2&15.5& 0.8&2&412&232&2&14.6& 0.5\\
& 200--300 & 6&0.41&0.18&6&14.3& 0.8&8&$<$100& 70&8&$<$14.0& 0.4\\
All Galaxies \\
& 0--100 & 11&0.68&0.64&11&15.3& 1.6&12&$<$146&127&12&$<$14.2& 0.5\\
& 100--300 & 26&0.28&0.23&26&14.0& 0.7&25&$<$ 62&107&26&$<$13.8& 0.4\\
& 300--500 & 32&0.16&0.14&31&13.7& 0.7&18&$<$ 44& 33&21&$<$13.6& 0.4\\
& 500--1000 & 39&0.05&0.16&38&13.0& 0.8&17&$<$ 40& 36&18&$<$13.6& 0.4\\
\enddata
\tablecomments{For column densities, all statistics are calculated on
  $\log N$. For the columns, $\mathcal{N}$ gives the number of
  galaxies, M() refers to median statistic, and R() refers to RMS.}
\tablenotetext{a}{Spectral type (see text for the quantitative definition).}
\tablenotetext{b}{Impact parameter and luminosity of the brightest galaxy ($L>0.1L_*$) within 200 kpc and $|\delta v| < 300 \mkms$ (if any).}
\end{deluxetable*}

Because the \lya\ line is saturated for most of the systems, the \wlya\ values
are a good proxy for kinematics, i.e.\ the systems at $\rho <
100$\,kpc require a velocity spread 
of $\Delta v = c \mwlya/(1215.67 \rm \AA) \approx 250 \, \mkms
(\mwlya/1\rm\AA)$.  Such motions may be just consistent with galaxies having
circular velocities
$v_c \gtrsim 100\mkms$, a reasonable estimate for \subls\ galaxies.
There is also the hint of a difference in \lya\ equivalent width for galaxies
with differing spectral type: the late-type galaxies have
systematically higher \wlya\ values than early-type galaxies at a
similar impact parameter.  This conclusion is tempered, however, by
the relatively small sample size and should be confirmed with a larger
dataset.

Remarkably, the sub-$L*$ galaxies show a very high incidence of
associated \ovi\ absorption for impact parameters $\rho < 300$\,kpc.  
At $\rho < 100$\,kpc, 3/4 (75\%) of the galaxies 
have
$\mwovi > 50$m\AA\ ($\mnovi > 10^{13.5} \cm{-2}$).  And at larger
impact parameters, the incidence is yet higher. 
It appears the association
extends well beyond the virial radii of these galaxies;
the results require that \subls\ galaxies are surrounded by a CGM that 
bears \ovi\ gas to an impact parameter of several hundreds kpc.
We also emphasize that only one of the galaxies has a neighboring, bright ($L>L^*$)
galaxy close to the sightline. 
Finally, we note a possible trend of
decreasing \wovi\ (and \novi) with increasing impact parameter. 
This trend, however, is dominated by the two
cases at $\rho \approx 100$\,kpc with large \wovi\ values.

The non-detection of \ovi\ absorption for the galaxy at $z=0.081$
and $\rho = 33$\,kpc in the FJ2155--0922 field warrants
additional discussion.  It is striking, given the high incidence
of \ovi\ detections, that the galaxy at smallest impact parameter
would exhibit the lowest \wovi\ value.\footnote{We have confirmed 
  the measurements of \cite{dsr+06} by analyzing our
  extraction of the {\it FUSE} spectrum for FJ2155--0922.}
Furthermore, as is evident from Figure~\ref{fig:subls} and Table~\ref{tab:subls},
there is very strong \lya\ absorption associated with this galaxy
($\mwlya > 1$\AA).  It has an early-type spectrum, and its
luminosity places it at the upper end of the \subls\ population.
It is noteworthy that this galaxy lies within an
overdensity in the FJ2155--0922 field: 10 galaxies with $L>0.1L^*$ lie within 3\,Mpc of the
sightline.  On the other hand, the $z=0.0788$, \subls\ galaxy $(\rho = 221$\,kpc)
in the same field has 22 such neighbors yet shows modest \ovi\ absorption.
Similar to the non-detection of \lya\ at $\rho \approx 30$\,kpc for
the dwarf galaxy in the PG1216+069 field, this one non-detection of \ovi\ gas
stresses the complexity of galaxy/absorber associations. 


In summary, all three classes of galaxies considered above exhibit
high covering fractions to \lya\ absorption for impact parameters
$\rho < 300$\,kpc.  There are possible trends of increasing \wlya\
values for brighter galaxies and decreasing \wlya\ values at higher
impact parameter.  In contrast, the covering fraction to \ovi\
absorption is highest for \subls\ galaxies, low for dwarf galaxies,
and low for early-type (and/or high-$\rho$) \lstar\ galaxies.  These
results are empirical descriptions of the gas associated with \lowz\
galaxies; we explore further the implications for the origins of \lya\
and \ovi\ absorption in the following sections.

\subsection{Galaxies Associated with IGM Absorption}
\label{sec:step2}

In the previous sub-section, we studied the \lya\ and \ovi\ absorption
associated with galaxies at low impact parameters to quasar
sightlines.  For that analysis, the measures
(luminosity, impact parameter, equivalent width) were all well-defined
and the results were relatively insensitive to biases related to
incompleteness in the galaxy survey.  The key
complication of such analysis, however, is that galaxies cluster and therefore their
environment (e.g.\ the presence of a brighter, nearby neighbor) may
play a significant role in observed associations between a galaxy
and the IGM.
Furthermore, this approach only indirectly addresses fundamental
questions related to the origin of \lya\ and \ovi\ gas in the
IGM.

One may gain further insight by reversing the
experiment, i.e.\ to study the properties of the closest (detected)
galaxy
to a given set of IGM absorption lines.  In this manner, one tests
how frequently galaxies and their halos are associated with
various components of the IGM, and one may assess their physical
properties and environment.
This approach, however, has the
obvious downside that no galaxy survey is truly complete, neither in
terms of field-of-view nor depth.  Indeed, our survey does not
cover sufficient area to identify all galaxies within $\rho \approx
200$\,kpc for $z<0.02$ and our magnitude limit implies significant
incompleteness to $L < 0.1L^*$ in most fields for $z>0.1$.  Therefore, we restrict the
following analysis and discussion to the redshift interval $0.02 < z <
0.2$ and caution that at the high end we have limited sensitivity to
very faint galaxies.

\begin{figure*}[ht]
\includegraphics[height=6.5in,angle=90]{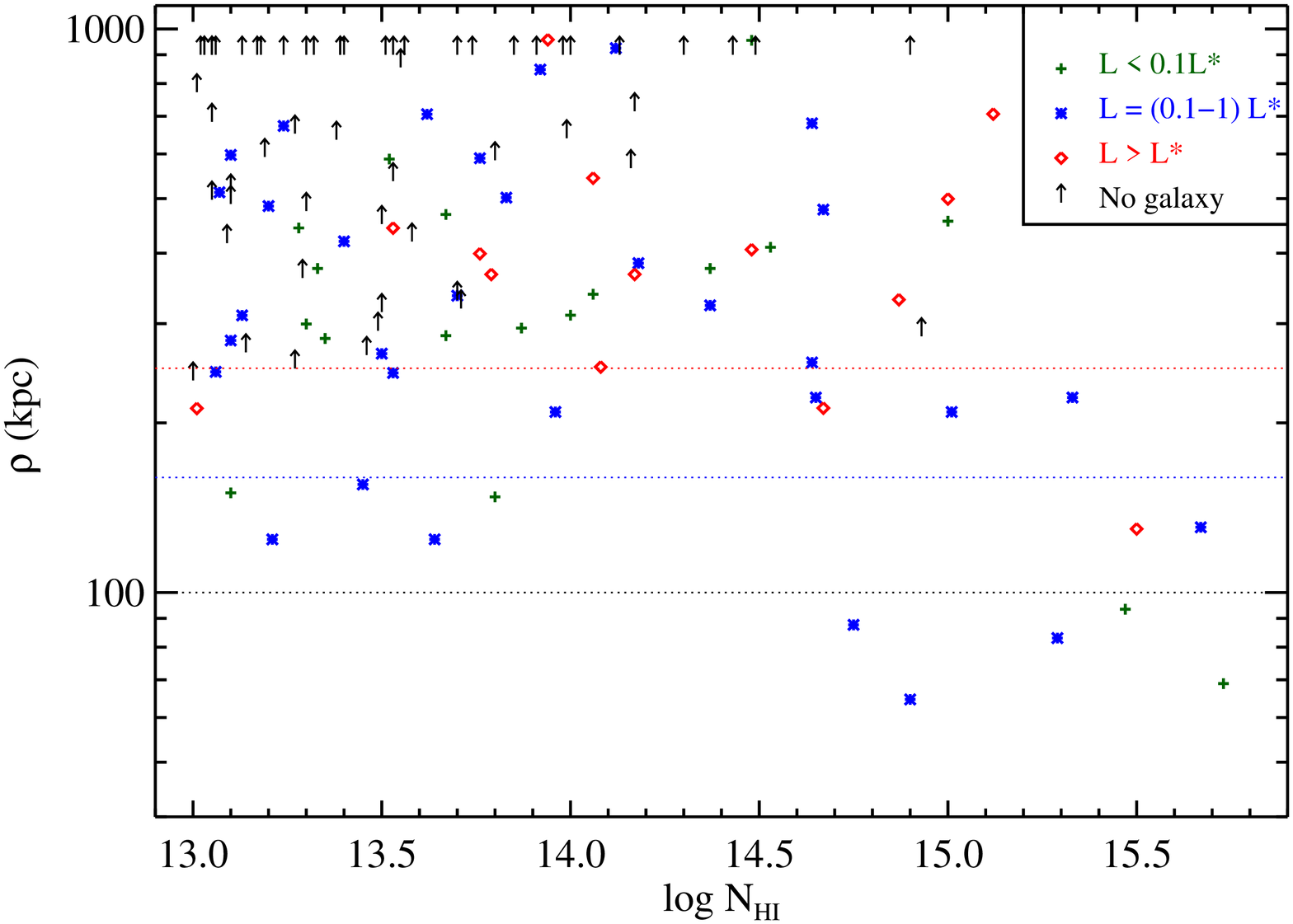}
\caption{Impact parameter (symbols indicate luminosity) of the
  galaxy closest to the sightline with a redshift coincident 
($|\delta v| < 400\,\mkms$) to the \lya\ absorbers detected along our quasar
sightlines. Note that most of the \nhi\ values exceeding $10^{14.5}
\cm{-2}$ should be considered lower limits due to line saturation. 
The dotted lines
  (from bottom to top) mark estimates for the virial radii of the
  dwarf, sub-$L*$, and $L*$ galaxies, respectively.   In cases where we
  have not detected a galaxy within 1\,Mpc of the sightline, we plot
  an upward arrow at the distance at each absorber's redshift
 corresponding to the typical $10'$
  field-of-view for our LCO/WFCCD galaxy survey. For $\mnhi < 10^{14.5}
  \cm{-2}$, the incidence of associated galaxies is low and those that
  are detected generally lie at  $\rho > 200$\,kpc.  We conclude that
  few, if any, of these absorbers arise in the virialized halos of $z
  \sim 0$ galaxies.  At higher column densities, the fraction of \lya\
  absorbers with an associated galaxy increases and the impact
  parameter to the systems decreases suggesting a physical association
  between the gas and galaxy. 
}
\label{fig:lya_gal}
\end{figure*}

In Figure~\ref{fig:lya_gal}, we present the analysis for \lya\ where
the symbols label the luminosity of the closest galaxy detected and
the upward arrows\footnote{These arrows are placed at the physical
  distance equivalent to $10'$ angular separation (the typical
  field-of-view for the LCO/WFCCD survey) at the redshift of the absorber.} 
indicate absorbers without an associated galaxy in
our survey. 
For very low column densities, $10^{13} \cm{-2} < \mnhi < 10^{14}
\cm{-2}$, approximately 40\%\ of the absorbers may be associated with
a galaxy within 1\,Mpc, always at $\rho > 100$\,kpc and primarily 
at $\rho > 200$\,kpc (80\% of the associations).
We conclude that {\it few, if any, of the weak \lya\ absorbers
arise from the virialized halos of $z \sim 0$ galaxies.}  
Instead, weak \ion{H}{1} absorption traces either the extended CGM of galaxies 
or, more commonly, structures with a low filling factor of galaxies (non-detections).  The
results are similar for the \lya\ lines with modest \nhi\
($10^{14}$--$10^{15} \cm{-2}$), although the incidence of an association
to a galaxy with $\rho < 1$\,Mpc is higher (70\%).  For the small set of strong
\ion{H}{1} absorption lines ($\mnhi > 10^{15} \cm{-2}$), the results are
qualitatively different.  Every \lya\ line is linked to a galaxy and
the majority of these have $\rho < 200$\,kpc.  Even in these cases,
however, the association with a virialized halo may be tenuous. 
Altogether, our results do offer further evidence that large \nhi\ absorbers
are physically associated with galaxies whereas weak \lya\ lines have
no direct physical connection \citep{cpw+05,shone+10}.

\begin{figure*}
\includegraphics[height=6.5in,angle=90]{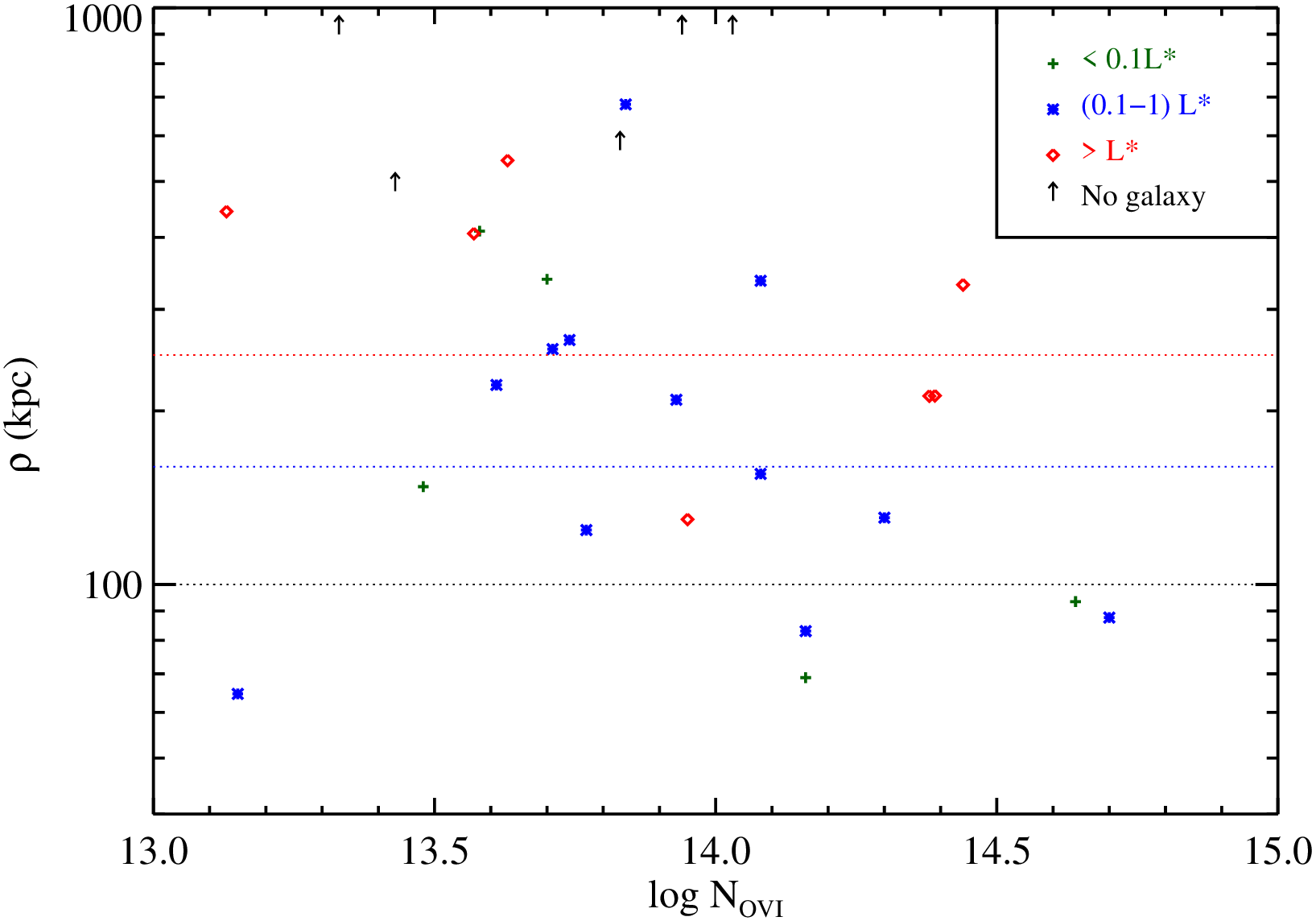}
\caption{
Impact parameter (symbols indicate luminosity) of the
  galaxy closest to the sightline with a redshift coincident 
($|\delta v| < 400\,\mkms$) to the \ovi\ absorbers detected along our
quasar sightlines.
The dotted horizontal lines mark estimates 
(from bottom to top) for the virial
radii of the dwarf, sub-$L*$, and $L*$ galaxies, respectively.
The arrows are positioned s in Figure~\ref{fig:lya_gal}.
One notes that although most of the \ovi\ absorbers have an associated
galaxy within a few hundred kpc of the sightline, very few of the
sightlines intersect the galaxy's expected virialized halo.
The results suggest that \ovi\ gas is associated primarily with the
extended CGM of $z \sim 0$ galaxies.  There is possibly an anti-correlation 
between $\rho$ and $N_{\rm OVI}$,  but the significance of such a trend is tempered by the
large scatter.
}
\label{fig:ovi_gal}
\end{figure*}

Turning to \ovi\ absorption,
Figure~\ref{fig:ovi_gal} shows the galaxy at the closest impact
parameter to the 30 published \ovi\ absorbers along our quasar
sightlines with $0.02 < z < 0.2$ (Table~\ref{tab:oviabs}).
This includes 5 \ovi\ systems where we have not identified a galaxy
within 1\,Mpc of the sightline with $|\delta v| < 400\mkms$.  
The rate of non-detections is much lower\footnote{ 
  This lends further support to the assertion that
  the lower fraction of weak \ion{H}{1} absorbers associated with galaxies
  is not driven by incompleteness in the LCO/WFCCD galaxy survey.}
than that observed for
weak \lya\ lines.  
Of course, only a few of the \ovi\ systems have
low \ion{H}{1} column densities ($\mnhi < 10^{14} \cm{-2}$).
Ignoring the non-detections, the median impact parameter to a galaxy is
approximately 200\,kpc \citep[consistent with the results
of][]{stockeetal06}.  This offset is smaller than that observed for
\lya\ lines having $\mnhi \approx 10^{14} \cm{-2}$.  On the other hand, only
a small fraction of the \ovi\ systems (5/30) have a galaxy detected
within 100\,kpc and only 1/18 for $\mnovi < 10^{14} \cm{-2}$.   
There are very few cases in the sample where one would insist that the
sightline has intersected the virialized halo of the associated
galaxy.  Instead, the \ovi\ gas appears linked to the extended CGM of
galaxies on scales of a few hundred kpc.
The luminosities of the nearest neighbors include galaxies at a wide
range of luminosity and spectral type, with a preference toward
late-type \subls\ systems.
There is an apparent trend of decreasing $\rho$ with increasing \novi\ for the range of
values considered in our analysis, but this is tempered by the large
scatter in these quantities.

As emphasized at the start of this sub-section, the results presented
in Figures~\ref{fig:lya_gal} and \ref{fig:ovi_gal} are subject
to the incompleteness of our galaxy survey.  In particular, one may argue
that a population of fainter galaxies lie at smaller impact parameters
than the sample comprising our survey.  
This could significantly affect our inferences
regarding the roles of virialized halos and the extended CGM in hosting
\lya/\ovi\ absorption.
To assess this concern,
Tables~\ref{tab:oviabs} lists the luminosity limit for each absorber
and the completeness of the galaxy survey to that limit for $\rho =
150$\,kpc.    Regarding completeness, with the exception of two fields
(PG1116+215 and PG1211+143), we obtained a redshift for every galaxy
within $\rho = 150$\,kpc of the \ovi\ absorber to the magnitude limit
$R = 19.5$\,mag.  We are confident, therefore, that no galaxy of
comparable luminosity lies at closer distance to the sightline than
the ones we have discovered.\footnote{Although we are not sensitive to
  galaxies at very small impact parameters to the quasar (i.e.\
  $\theta \lesssim 1''$, or $\rho \lesssim 20$\,kpc for the redshifts
  considered), we expect these to be very rare occurrences
  and note that none of the sightlines exhibit very large \ion{H}{1} column
  densities ($\mnhi \gg 10^{17} \cm{-2}$).}
Regarding the luminosity limit, a subset of the galaxies lie within
twice the limit yet $\approx 40 \%$ are $5 \times$ brighter than
the formal limit.  Furthermore, owing to the details of mask design
and specific scientific interests,  we have surveyed several fields to
$R>19.5$\,mag albeit with lower completeness.  The galaxies discovered
to these fainter magnitudes have been included in the analysis.
We conclude, therefore, that our results are unlikely to be
severely compromised by a lack of sensitivity to fainter galaxies.

\begin{deluxetable*}{ccccccccccc}
\tablewidth{0pc}
\tablecaption{\ovi\ Absorbers Identified in our Fields\label{tab:oviabs}}
\tabletypesize{\scriptsize}
\tablehead{\colhead{Field} & \colhead{$z_{\rm OVI}$} &
\colhead{$W^{\rm OVI}$} & \colhead{$N_{\rm OVI}$} & \colhead{Ref} &
\colhead{$z_{\rm n}^a$} & \colhead{$\rho_{\rm n}^a$} & \colhead{$L_{\rm n}^a$} &
\colhead{$L_{\rm lim}^b$} & \colhead{$C^c$} \\
 & & (m\AA) & & & & (kpc) & ($L^*$) & ($L^*$) }
\startdata
\cutinhead{$0.005 < z < 0.02$}
PG1116+215 & 0.0050 & \ldots &13.91 & 4 & \dots & \dots & \dots & 0.000 &  66\\
PG1211+143 & 0.0071 & \ldots &13.78 & 3 & \dots & \dots & \dots & 0.000 &  52\\
3C273 & 0.0033 & $  31$& 13.41 & 1 & \dots & \dots & \dots & 0.000 &  73\\
3C273 & 0.0076 & $  14$& 13.11 & 3 & \dots & \dots & \dots & 0.001 &  74\\
FJ2155-0922 & 0.0173 & $< 350$& 13.90 & 4 & \dots & \dots & \dots & 0.003 & 100\\
\cutinhead{$0.02 < z < 0.2$}
TONS180 & 0.0456 & $  62$& 13.74 & 4 & 0.0456 &  265 & 0.377 & 0.020 & 100\\
TONS180 & 0.0234 & $  43$& 13.48 & 4 & 0.0234 &  148 & 0.019 & 0.005 & 100\\
TONS180 & 0.0436 & $  31$& 13.43 & 4 & \dots & \dots & \dots & 0.018 & 100\\
PKS0312-77 & 0.1983 & $  64$& 13.84 & 1 & 0.1982 &  680 & 0.785 & 0.464 & 100\\
PKS0312-77 & 0.1589 & $  86$& 13.94 & 1 & \dots & \dots & \dots & 0.284 & 100\\
PKS0405-123 & 0.0918 & $  49$& 13.58 & 4 & 0.0906 &  410 & 0.067 & 0.087 & 100\\
PKS0405-123 & 0.0966 & $  59$& 13.71 & 3 & 0.0965 &  256 & 0.150 & 0.097 & 100\\
PKS0405-123 & 0.1669 & $ 412$& 14.64 & 1 & 0.1670 &   93 & 0.080 & 0.316 & 100\\
PKS0405-123 & 0.1829 & $  84$& 13.94 & 1 & \dots & \dots & \dots & 0.387 & 100\\
PG1116+215 & 0.1655 & $ 111$& 14.08 & 1 & 0.1660 &  156 & 0.336 & 0.311 &  80\\
PG1116+215 & 0.1385 & $  83$& 13.95 & 1 & 0.1383 &  130 & 2.039 & 0.210 &  80\\
PG1116+215 & 0.0593 & $  63$& 13.77 & 1 & 0.0600 &  124 & 0.104 & 0.035 &  67\\
PG1211+143 & 0.0513 & \ldots &14.30 & 4 & 0.0511 &  131 & 0.829 & 0.026 &  64\\
PG1211+143 & 0.0645 & $ 144$& 14.16 & 3 & 0.0646 &   69 & 0.084 & 0.041 &  75\\
PG1216+069 & 0.1236 & $ 378$& 14.70 & 1 & 0.1241 &   88 & 0.650 & 0.164 & 100\\
3C273 & 0.0902 & $  16$& 13.13 & 1 & 0.0902 &  443 & 1.105 & 0.084 & 100\\
3C273 & 0.1200 & $  24$& 13.33 & 1 & \dots & \dots & \dots & 0.154 & 100\\
PKS1302-102 & 0.0940 & $  19$& 13.15 & 4 & 0.0936 &   65 & 0.195 & 0.091 & 100\\
PKS1302-102 & 0.0647 & $  69$& 13.70 & 3 & 0.0647 &  338 & 0.049 & 0.042 & 100\\
PKS1302-102 & 0.0989 & \ldots &14.03 & 3 & \dots & \dots & \dots & 0.102 & 100\\
PKS1302-102 & 0.0423 & $ 194$& 14.38 & 3 & 0.0426 &  212 & 2.577 & 0.017 &  80\\
PKS1302-102 & 0.1453 & $ 146$& 14.16 & 4 & 0.1453 &   83 & 0.396 & 0.233 & 100\\
PKS1302-102 & 0.1916 & $  80$& 13.93 & 1 & 0.1917 &  209 & 0.888 & 0.430 & 100\\
MRK1383 & 0.0519 & $  66$& 13.83 & 4 & \dots & \dots & \dots & 0.026 &  60\\
FJ2155-0922 & 0.1579 & $ 120$& 14.08 & 1 & 0.1581 &  336 & 0.495 & 0.280 & 100\\
FJ2155-0922 & 0.1324 & $ 225$& 14.39 & 1 & 0.1326 &  212 & 1.626 & 0.191 & 100\\
FJ2155-0922 & 0.0776 & $  46$& 13.61 & 1 & 0.0788 &  222 & 0.157 & 0.061 & 100\\
FJ2155-0922 & 0.1765 & \ldots &14.44 & 1 & 0.1764 &  331 & 2.399 & 0.358 & 100\\
PKS2155-304 & 0.0571 & $  44$& 13.57 & 3 & 0.0570 &  406 & 1.595 & 0.032 & 100\\
PKS2155-304 & 0.0541 & $  32$& 13.63 & 3 & 0.0541 &  544 & 2.053 & 0.029 & 100\\
\cutinhead{$z>0.2$}
PKS0312-77 & 0.2027 & $ 538$& 14.91 & 1 & 0.2026 &  127 & 0.634 & 0.487 & 100\\
PKS0405-123 & 0.4950 & $ 213$& 14.44 & 1 & \dots & \dots & \dots & 3.855 & 100\\
PKS0405-123 & 0.3509 & $  24$& 13.35 & 3 & 0.3520 &  157 & 0.273 & 1.707 & 100\\
PKS0405-123 & 0.3616 & $  96$& 14.00 & 1 & 0.3612 &  214 & 6.604 & 1.831 & 100\\
PKS0405-123 & 0.3634 & $  38$& 13.61 & 1 & \dots & \dots & \dots & 1.852 & 100\\
PG1216+069 & 0.2677 & $  22$& 13.30 & 1 & \dots & \dots & \dots & 0.913 & 100\\
PG1216+069 & 0.2819 & $  94$& 13.95 & 1 & \dots & \dots & \dots & 1.028 & 100\\
PKS1302-102 & 0.2274 & $  40$& 13.59 & 1 & \dots & \dots & \dots & 0.631 & 100\\
PKS1302-102 & 0.2256 & $  82$& 13.93 & 1 & 0.2256 &  417 & 0.892 & 0.620 & 100\\
PKS1302-102 & 0.2044 & $  33$& 13.64 & 3 & \dots & \dots & \dots & 0.496 & 100\\
PKS1302-102 & 0.2529 & $  11$& 12.92 & 3 & \dots & \dots & \dots & 0.802 & 100\\
\enddata
\tablecomments{The analysis presented in $\S$~\ref{sec:step2} only considers \ovi\ systems with $0.02 < z < 0.2$ to minimize incompleteness in the galaxy survey.}
\tablenotetext{a}{Impact parameter and luminosity of the brightest galaxy ($L>0.1L^*$) within 200 kpc and $|\delta v| < 300 \mkms$ (if any).}
\tablenotetext{b}{Impact parameter and luminosity of the brightest galaxy ($L>0.1L^*$) within 200 kpc and $|\delta v| < 300 \mkms$ (if any).}
\tablenotetext{c}{Completeness of the galaxy survey to $R=19.5$\,mag\ within the radius corresponding to $\rho = 150$\,kpc at the \ovi\ absorber's redshift.}
\end{deluxetable*}

Another (brute force) approach to assess the effects of incompleteness
is to obtain significantly deeper spectroscopy in the fields.
\cite{cm09} have performed a galaxy survey with the IMACS and LDSS3
spectrometers on the 6.5\,m Magellan telescopes for two of our fields
(PKS0405--123, PG1216+069).  They achieved a high completeness around
each quasar sightline to
several magnitudes fainter limits ($\approx 10\times$ deeper).  For the
PKS0405--123 field, they do identify a new dwarf galaxy associated
with the $z=0.0918$ \ovi\ absorber which lies at closer impact
parameter ($\rho = 73$\,kpc) than the dwarf galaxy we had discussed.
This is, however, the only modification for the four \ovi\ absorbers
within the redshift interval $0.02 < z < 0.2$, and even this galaxy 
is at an impact parameter that may exceed its virialized halo. 
For the PG1216+069 field, \cite{cm09} find no fainter galaxies at
closer impact parameters for the $z=0.125$ \subls\ galaxy 
identified in our survey.

\section{Discussion}
\label{sec:discussion}

The previous section described the association of \lya\ and \ovi\
absorption with galaxies discovered in our LCO/WFCCD survey
(Paper~IV).  In $\S$~\ref{sec:step1}, we explored the strength of
\lya\ and \ovi\ absorption as a function of galaxy luminosity,
spectral type, and impact parameter.  We then reported on the results
of a search for galaxies related to \lya\ and \ovi\ absorption
($\S$~\ref{sec:step2}).  In this section, we synthesize these results
and discuss the implications in the context of previous observational
and theoretical work.  We divide the discussion by transition (i.e.\
\lya\ and \ovi) but emphasize that a complete description of the IGM
must consider both elements.

\subsection{\lya}
\label{sec:discuss_lya}


The \lya\ transition is the strongest and most commonly observed
transition of the intergalactic medium.  Its complex pattern of
absorption defines the so-called \lya\ forest and sets the starting
point for nearly all IGM analysis.  The current cosmological paradigm
for the \lya\ forest is that these lines arise from an undulating
Gunn-Peterson field of overdensities \citep[e.g.][]{mco+96}.  In turn,
observations of the IGM provide direct constraints for cosmological
parameters \citep[e.g.][]{wmap05}.  
This gas is also predicted to be the dominant baryonic reservoir of
the universe, which fuels the formation and growth of galaxies.

We begin with a discussion of the properties of \ion{H}{1} gas
surrounding low-$z$ galaxies ($\rho < 300$\,kpc).  Several previous
studies have examined the incidence
of \lya\ absorption for galaxies located at small impact parameters
to quasar sightlines.  In each case, the authors have reported a
very high detection rate for galaxies within a few hundred kpc: 
\cite{clw01} found positive detections for
29/31 of their predominantly \lstar\ galaxies to $\rho = 250\,
h^{-1}_{72}$\,kpc with $\mwlya > 300$\,m\AA\ and reported a sharp
decline at larger impact parameter; \cite{bowen+02} detected
\lya\ lines with $\mnhi \ge 10^{13} \cm{-2}$ for 8/8 of the galaxies
located within $\rho = 280 \, h_{72}^{-1}$\,kpc of their quasar
sightlines,
and \cite{wakker09} reported positive detections for 7/7 of the $L \ge 0.1
L^*$ field galaxies
in their sample when restricting to $\rho \le 350$\,kpc.  

With our LCO/WFCCD survey we have examined a sample of 37 galaxies
at $\rho < 300$\,kpc
with greater emphasis on lower luminosities than previous work.
The results presented in Figure~\ref{fig:dwarfs}--\ref{fig:subls}
(Tables~\ref{tab:dwarfs}--\ref{tab:subls}) further solidify the
correlation
between galaxies and \lya\ absorption: to an impact
parameter of $\rho \approx 300$\,kpc, we observe a nearly 100\%\
probability for detecting \lya\ absorption ($\mwlya > 50$\,m\AA) 
within $400 \mkms$ of a
galaxy.   This result is independent of galaxy luminosity (for
$L>0.01L^*$), spectral type, and local environment\footnote{ 
  Our results do not confirm the findings of \cite{wakker09} that galaxies in
  groups exhibit systematically weaker \lya\ absorption, although we
  have not yet constructed a well-defined group sample from the
  LCO/WFCCD survey.}
(defined by the number of close neighbors).  
Altogether, these results demand that low-$z$ galaxies of all types
are embedded within
a diffuse, highly ionized medium with a very high covering fraction
to significant \ion{H}{1} absorption. 
We refer to this medium as the circumgalactic medium (CGM),
although this gas need not be physically associated (e.g.\ bound) to
the galaxy.

Before considering the origin of this medium, we review and then
explore its physical properties.  First, we note that the Doppler
parameters that are typical of low-$z$ \lya\ lines generally require kinetic
temperatures $T < 10^5$\,K \citep[e.g.][]{dhk+99,willigeretal06}.  The
standard expectation is that
the majority of these absorbers trace photoionized
gas with $T \sim 10^4$\,K.  For those absorption systems that also exhibit
metal-line transitions (e.g.\ \ion{C}{4}, \ion{C}{3}, \ion{O}{6}),
this hypothesis may be tested against ionization modeling
and the results are usually consistent with a photoionized medium
\citep[e.g.][]{pks0405_uv,cpc+08}.  This modeling also provides a
constraint on the ionization fraction and an estimate of the total
hydrogen column density $N_H$.  

A remarkable result for the  low-$z$
IGM is that the estimated $N_{\rm H}$ values are
uniform at $\approx 10^{19} \cm{-2}$, i.e., independent of the observed \nhi\ value
\citep{pks0405_uv,lsr+07}.  This `$N_{\rm H}$ constancy' was interpreted by
\cite{pks0405_uv} to result from
lower \nhi\ absorbers having lower average volume density
\citep[e.g.][]{dhk+99} which implies a
higher ionization correction for a uniform (background) radiation field.
Indeed, analysis of recent simulations of the low-$z$ IGM have
considered the $N_{\rm H}$ constancy explicitly and have reproduced the
observed result with a similar explanation \citep{dave+10}.
We caution
that this $N_{\rm H}$ constancy was derived empirically from absorbers with
$\mnhi \approx 10^{13.5} -- 10^{15} \cm{-2}$ which had no known
association to galaxies.  We demonstrate below, however, that the
majority of the stronger \nhi\ systems must be associated with
galaxies.

Adopting a constant $N_{\rm H}$ value for the observed \lya\ forest, we may 
crudely estimate the mass of the photoionized CGM surrounding
\lowz\ galaxies.  A simple estimate follows from the characteristic hydrogen column density
$\mncgm$ and a characteristic radius \rcgm:

\begin{equation}
M_{\rm CGM} = \mncgm m_p \mu \, \pi r_{\rm CGM}^2 \cmma
\end{equation}
with the factor $\mu \approx 1.3$ accounting for helium.
For our assumed values, we find

\begin{equation}
M_{\rm CGM} \approx 3 \sci{10} \, M_\odot \; \ltp \frac{\mncgm}{10^{19}
\cm{-2}} \rtp \; \ltp \frac{\mrcgm}{300 \rm kpc} \rtp^2 \perd
\label{eqn:mass_CGM}
\end{equation}
Therefore, we estimate a baryonic reservoir surrounding each galaxy
which meets or significantly exceeds the typical baryonic mass of
present-day \subls\ galaxies.  Again, this CGM need not be physically associated
(e.g.\ gravitationally bound) with
the galaxy.  Nevertheless, it represents a large baryonic
reservoir for future star formation or could even (in part) be gas that was
expelled during the processes of galaxy formation.  
Lastly, if the gas is metal enriched, the mass in metals could exceed the
amount metals locked in stars within these galaxies.

\begin{figure}
\includegraphics[width=3.5in]{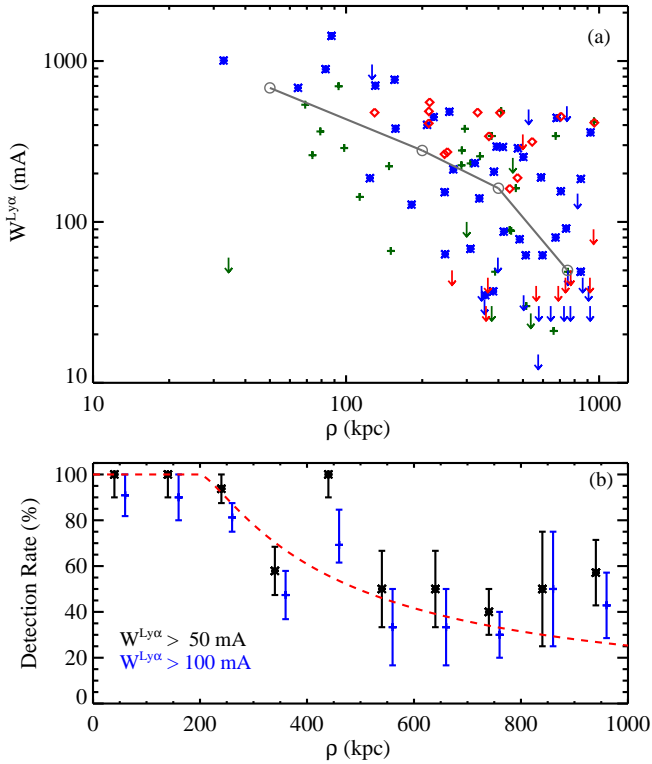}
\caption{(a) Measurements of associated \lya\ absorption for galaxies of
  all luminosity in our sample within 1\,Mpc of a quasar sightline
  with UV spectral coverage of the \lya\ line.  We have grouped
  galaxies together in intervals of 400\,\kms\ and only plot the system
  nearest to the sightline.  Red diamonds indicate \lstar\ galaxies,
  blue asterisks are \subls\ galaxies, and green plus-signs indicate
  dwarf galaxies.  Arrows of the same colors represent non-detections
  for galaxies of these luminosities.  There is a strong anti-correlation
  between \lya\ equivalent width (and column density) and impact
  parameter.  Furthermore, there is a very high incidence of positive
  detections to 50\,m\AA\ to $\rho \approx 500$\,kpc and a
  marked decline at larger impact parameter.  The gray circles and
  line trace the median equivalent width (including limits) in bins 
  of $\rho = \rm [0--100]\,kpc, [100--300]\,kpc, [300--500]\,kpc, [500--1000]\,kpc$
  (Table~\ref{tab:gastat}).
  (b) Detection rate of a \lya\ line within 400\,\kms\ of the galaxies
  in the LCO/WFCCD survey as a function of impact parameter, for two
  limiting equivalent widths.  Error bars reflect a bootstrap analysis
  which provide a crude estimate of the uncertainty in the
  measurements.  The dashed red line indicates the prediction for the
  detection rate for a simplistic filamentary model with filamentary
  width $w \approx 400$\,kpc (see text for details).
}
\label{fig:1Mpc}
\end{figure}

\begin{deluxetable*}{crrccccccccccc}
\tablewidth{0pc}
\tablecaption{Galaxies within 1\,Mpc of a QSO Sightline\label{tab:1Mpc}}
\tabletypesize{\scriptsize}
\tablehead{\colhead{Field} & \colhead{RA} & \colhead{DEC} & \colhead{$z_{\rm gal}$} &  \colhead{$\rho$} & 
\colhead{$L$}  & \colhead{$z_{\rm abs}^{\mlya}$} & \colhead{$W^{\mlya}$} &
\colhead{\nhi} & \colhead{Ref.} & \colhead{$z_{\rm abs}^{\rm OVI}$} &
\colhead{$W^{\rm OVI}$} & \colhead{$N_{\rm OVI}$} & \colhead{Ref} \\
 & (J2000) & (J2000) & & (kpc) & ($L_{*}$) & & (m\AA) & & & & (m\AA) }
\startdata
Q0026+1259 & 00:29:09.32 & +13:16:28.5 & 0.0329 &   43 &  0.018 & \dots & \dots & \dots & \dots &0.0329 & $< 120$&$<14.00$& 9\\
  & 00:29:15.36 & +13:20:57.0 & 0.0393 &  213 &  1.384 & \dots & \dots & \dots & \dots &0.0393 & $< 120$&$<14.00$& 9\\
  & 00:28:53.55 & +13:24:19.9 & 0.0565 &  594 &  0.559 & \dots & \dots & \dots & \dots &\dots & \dots & \dots & \dots \\
  & 00:29:53.44 & +13:10:49.5 & 0.0738 &  880 &  0.186 & \dots & \dots & \dots & \dots &\dots & \dots & \dots & \dots \\
  & 00:29:16.48 & +13:21:52.9 & 0.0804 &  497 &  0.113 & \dots & \dots & \dots & \dots &\dots & \dots & \dots & \dots \\
  & 00:29:37.55 & +13:10:21.0 & 0.0961 &  822 &  1.958 & \dots & \dots & \dots & \dots &0.0961 & $< 100$&$<14.00$& 9\\
  & 00:29:23.49 & +13:09:40.8 & 0.1125 &  782 &  1.091 & \dots & \dots & \dots & \dots &0.1125 & $< 100$&$<14.00$& 9\\
  & 00:29:16.04 & +13:10:02.5 & 0.1314 &  792 &  0.179 & \dots & \dots & \dots & \dots &\dots & \dots & \dots & \dots \\
TONS180 & 00:57:04.01 & --22:26:51.2 & 0.0234 &  147 &  0.019 & 0.0234 &$ 222$&$13.80$& 5 & 0.0234 & $  43$&$13.48$& 4\\
  & 00:57:08.52 & --22:18:29.6 & 0.0456 &  265 &  0.377 & 0.0456 &$ 212$&$13.80$& 5 & 0.0456 & $  62$&$13.74$& 4\\
\enddata
\tablecomments{Galaxies within $\pm 400\mkms$ in redshift were grouped together and only the member with smallest impact parameter is tabulated here.}
\tablecomments{[The complete version of this table is in the electronic edition of the Journal.  The printed edition contains only a sample.]}
\end{deluxetable*}

The near unity covering fraction of this CGM for \lya\ absorption to
$\rho \approx 300$\,kpc inspires us to examine further its radial
extent.  Previously, \cite{chenetal01} reported a decline in the
average \lya\ equivalent width for $\rho \gtrsim 250$\,kpc.
Similarly, \cite{tripp+98} and \cite{wakker09} noted a
lower detection rate and average \lya\ equivalent width for $L \approx
L^*$ galaxies to $\rho \gtrsim 1$\,Mpc.  
In Figure~\ref{fig:1Mpc} (Table~\ref{tab:1Mpc}), 
we have extended the analysis of our LCO/WFCCD survey to $\rho =
1$\,Mpc.
Here we include galaxies of all luminosity and plot only 
the closest galaxy to the sightline 
without repeating any pair of galaxies with $|\delta v| < 400 \mkms$.
We also include the galaxy survey in the field of PKS0405--123
performed by \cite{willigeretal06}.
There are two issues to emphasize before proceeding: (i)
incompleteness in our galaxy survey implies that the $\rho$ values are
strictly upper limits, i.e.\ another (probably fainter) galaxy may exist
at smaller impact parameter; and (ii) there may be little 
physical connection (e.g.\ gravitational interaction) between
this gas and the galaxy on such large scales.  Nevertheless,  the
results may offer a valuable constraint for galaxy/IGM models and 
provide further insight into the galaxy/absorber connection.

There are two obvious conclusions to draw from
Figure~\ref{fig:1Mpc}: 
(a) the median equivalent width decreases with increasing $\rho$; 
and 
(b) the detection rate declines beyond $\rho \approx 300$\,kpc but
apparently remains above the random
rate ($\approx 10\%$; $\S$~\ref{sec:step1}) out to at least 1\,Mpc.
These conclusions are not independent, e.g.\ the median \wlya\ value is
significantly lower at $\rho > 500$\,kpc because of the lower
detection rate.
A Spearman's correlation test of only the {\it detections} reports the
null hypothesis of no correlation is ruled out at $>99.99\%$\,c.l.
This anti-correlation is only strengthened by the non-detections, and
it is not a natural
consequence of incompleteness in the galaxy survey. 
We have performed a simple linear regression on the positive
detections in log-log space, i.e.\ by assuming $\mwlya(\rho) = W_0 (\rho/1\,\rm
kpc)^{-\gamma}$.  
If we ignore the \lstar\ galaxies (for reasons discussed below) and perform a linear
regression on the remaining galaxies with detections, 
we recover $W_0 = 4.0$\AA\ and $\gamma = -0.69$.  These values are in
good agreement with \cite{tripp+98}, whose analysis was performed on a
heterogenous set of galaxy samples.   We also note that the 
\wlya\ values have a shallower trend with impact parameter
for $\rho > 100$\,kpc.  Removing those galaxy/absorber pairs from the
linear regression, we recover $W_0 = 3.3$\AA\ and $\gamma = -0.43$.
These values may better describe evolution in the large-scale
structures that surround low-$z$ galaxies.
We also stress that a rank-correlation test on the detections
and upper limits for galaxies at $\rho > 200$\,kpc rules out the null
hypothesis at $>99.93\%$\,c.l.
Both the detection rate and median equivalent width of \lya\
absorption are correlated with galaxy impact parameter to at least
1\,Mpc, i.e.\ far beyond the virialized halos and presumed CGM of
these galaxies.  
Before concluding, we note
that a power-law description for
$\mwlya(\rho)$ was not physically motivated. 
Furthermore, despite the significant anti-correlation, it
is not a very good description of the data;  
there is tremendous scatter about the regression at all impact parameters.

\begin{figure}
\includegraphics[width=3.5in]{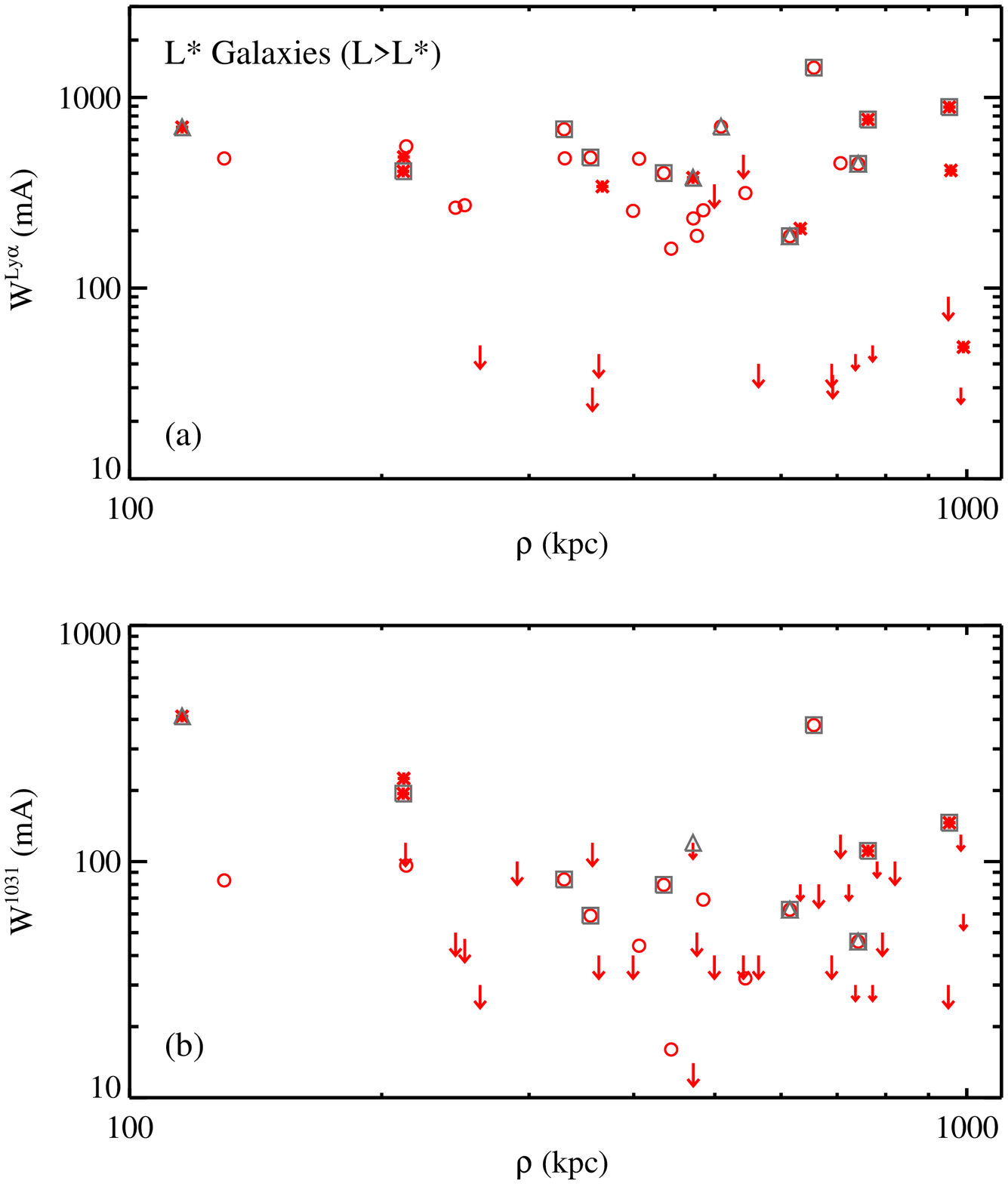}
\caption{(a) Measured \lya\ equivalent widths for every \lstar\ galaxy
  in the LCO/WFCCD survey (with the closest galaxy shown for cases
  with two or more within 400\,\kms) as a function of impact parameter,
  extending to 1\,Mpc.  
Galaxies with early-type spectra are marked with open circles and
those with late-type spectra are plotted as asterisks.  Small/large
arrows denote limits for late/early-type spectra.
The squares (triangles) indicate cases where an
  additional \subls\ (dwarf) galaxy with consistent redshift lies
  closer to the sightline.  The galaxies exhibit two sets of \wlya\
  measurements: 
  (i) positive detections with $\mwlya > 200$\,m\AA\ and
  (ii) a population of non-detections with $\mwlya < 100$\,m\AA\
  (generally $<50$\,m\AA).  
  (b)  Equivalent width of \ovi~1031 as a function of impact parameter
  for \lstar\ galaxies from the LCO/WFCCD survey.  Symbols have the
  same meaning as in (a).  Only a small fraction of these galaxies
  exhibit a positive \ovi\ detection for $\rho > 250$\,kpc, especially
  if one excludes the systems where a closer, fainter galaxy exists at
  the same redshift.  These results indicate that \lstar\ galaxies are
  not surrounded by a CGM to many hundred kpc with a high covering
  fraction to significant \ovi\ absorption.
}
\label{fig:lstar_1mpc}
\end{figure}

\begin{deluxetable*}{crrccccccccccccc}
\tablewidth{0pc}
\tablecaption{$L>L^*$ Galaxies within 1\,Mpc of a QSO Sightline\label{tab:lstar_1Mpc}}
\tabletypesize{\scriptsize}
\tablehead{\colhead{Field} & \colhead{RA} & \colhead{DEC} & \colhead{$z_{\rm gal}$} & \colhead{T$^a$} & \colhead{$\rho$} & 
\colhead{$L$} & \colhead{$\mathcal{N}^b$} & \colhead{$z_{\rm abs}^{\mlya}$} & \colhead{$W^{\mlya}$} &
\colhead{\nhi} & \colhead{Rf.} & \colhead{$z_{\rm abs}^{\rm OVI}$} &
\colhead{$W^{\rm OVI}$} & \colhead{$N_{\rm OVI}$} & \colhead{Rf.} \\
 & (J2000) & (J2000) & & & (kpc) & ($L^{*}$) & & & (m\AA) & & & & (m\AA) }
\startdata
Q0026+1259 & 00:29:15.36 & +13:20:57.0 & 0.0393 & E &  213 &  1.384 &  0 & \dots & \dots & \dots & &0.0393 & $< 120$&$<14.00$& 9\\
  & 00:29:37.55 & +13:10:21.0 & 0.0961 & E&  822 &  1.958 &  0 & \dots & \dots & \dots & &0.0961 & $< 100$&$<14.00$& 9\\
  & 00:29:23.49 & +13:09:40.8 & 0.1125 & L  &  782 &  1.091 &  0 & \dots & \dots & \dots & &0.1125 & $< 100$&$<14.00$& 9\\
PKS0312-77 & 03:12:31.11 & --76:43:25.0 & 0.0514 & E&  692 &  4.904 &  3 & \dots & \dots & \dots & &\dots & \dots & \dots & \dots \\
  & 03:11:02.46 & --77:00:05.2 & 0.0529 & E&  897 &  1.477 &  3 & \dots & \dots & \dots & &\dots & \dots & \dots & \\
  & 03:11:58.58 & --76:48:55.4 & 0.1192 & L &  366 &  1.846 &  1 & 0.1183 &$ 341$&$14.17$& 3 & \dots & \dots & \dots & \\
PKS0405-123 & 04:08:06.63 & --12:12:50.9 & 0.0800 & E&  399 &  1.404 &  4 & 0.0814 &$ 254$&$13.76$& 3 & 0.0800 & $<  40$&$<13.50$& 9\\
  & 04:07:54.22 & --12:14:50.7 & 0.0967 & E&  355 &  1.357 &  9 & 0.0966 &$ 484$&$14.64$& 3 & 0.0966 & $  59$&$13.71$& 3\\
  & 04:07:51.28 & --12:11:38.3 & 0.1670 & L &  115 &  2.143 &  1 & 0.1671 &$ 697$&$15.47$& 3 & 0.1669 & $ 412$&$14.64$& 1\\
  & 04:07:42.79 & --12:11:33.1 & 0.2030 & E&  262 &  1.070 &  9 & 0.2030 &$<  50$&$<13.00$& 9 & 0.2030 & $<  30$&$<13.40$& 9\\
  & 04:08:01.90 & --12:11:40.1 & 0.2484 & L &  736 &  1.289 &  1 & 0.2484 &$<  45$&$<12.90$& 9 & 0.2484 & $<  30$&$<13.40$& 9\\
  & 04:07:34.68 & --12:13:22.6 & 0.2951 & E&  950 &  1.198 &  3 & 0.2951 &$<  90$&$<13.20$& 9 & 0.2951 & $<  30$&$<13.40$& 9\\
  & 04:07:50.69 & --12:12:25.2 & 0.2976 & E&  245 &  1.878 &  7 & 0.2977 &$ 264$&$13.85$& 3 & 0.2976 & $<  50$&$<13.60$& 9\\
  & 04:07:52.33 & --12:14:08.0 & 0.3099 & E&  689 &  1.791 &  3 & 0.3099 &$<  40$&$<12.90$& 9 & 0.3099 & $<  40$&$<13.50$& 9\\
  & 04:07:57.96 & --12:09:52.5 & 0.3203 & L &  771 &  1.038 &  1 & 0.3203 &$<  50$&$<13.00$& 9 & 0.3203 & $<  30$&$<13.40$& 9\\
  & 04:07:50.24 & --12:09:52.2 & 0.3252 & E&  475 &  2.317 &  0 & 0.3250 &$ 188$&$13.81$& 3 & 0.3252 & $<  50$&$<13.60$& 9\\
  & 04:07:45.96 & --12:11:09.9 & 0.3612 & E&  213 &  6.604 &  5 & 0.3608 &$ 554$&$15.14$& 3 & 0.3616 & $  96$&$14.00$& 1\\
  & 04:07:51.88 & --12:13:16.6 & 0.4253 & E&  585 &  6.161 &  5 & \dots & \dots & \dots & &\dots & \dots & \dots & \\
  & 04:07:55.50 & --12:10:37.0 & 0.4282 & U  &  638 &  1.086 &  2 & \dots & \dots & \dots & &\dots & \dots & \dots & \\
  & 04:07:41.00 & --12:13:15.4 & 0.5563 & U &  893 &  1.854 &  0 & \dots & \dots & \dots & &\dots & \dots & \dots & \\
PG1004+130 & 10:07:34.55 & +12:52:09.5 & 0.0704 & E&  290 &  2.404 &  0 & \dots & \dots & & \dots &0.0704 & $< 100$&$<13.90$& 9\\
  & 10:07:25.39 & +12:53:04.8 & 0.1674 & E&  665 &  3.577 &  2 & \dots & \dots & \dots & &0.1674 & $<  80$&$<13.80$& 9\\
  & 10:07:10.67 & +12:50:04.3 & 0.1931 & L &  722 &  1.339 &  3 & \dots & \dots & \dots & &0.1931 & $<  80$&$<13.80$& 9\\
PG1116+215 & 11:19:24.31 & +21:10:30.6 & 0.0590 & E&  614 &  1.419 &  3 & 0.0593 &$ 187$&$13.64$& 1 & 0.0593 & $  63$&$13.77$& 1\\
  & 11:19:06.73 & +21:18:29.3 & 0.1383 & E&  129 &  2.039 &  4 & 0.1385 &$ 479$&$>14.35$& 1 & 0.1385 & $  83$&$13.95$& 1\\
  & 11:19:16.88 & +21:14:57.1 & 0.1652 & L &  761 &  1.150 & 14 & 0.1655 &$ 765$&$>14.43$& 1 & 0.1655 & $ 111$&$14.08$& 1\\
PG1211+143 & 12:14:01.15 & +14:11:07.2 & 0.0522 & E&  508 &  1.210 &  6 & 0.0510 &$ 703$&$15.67$& 3 & 0.0513 & \ldots&$14.30$& 4\\
PG1216+069 & 12:19:30.89 & +06:43:34.6 & 0.0805 & L &  470 &  2.252 &  1 & 0.0805 &$ 379$&$13.87$& 3 & 0.0815 & $< 120$&$<14.00$& 9\\
  & 12:19:40.12 & +06:32:01.5 & 0.1191 & L &  983 &  2.326 &  3 & 0.1184 &$<  30$&$<12.70$& 9 & 0.1191 & $< 130$&$<14.10$& 9\\
  & 12:19:33.88 & +06:42:44.2 & 0.1250 & E&  656 &  1.789 &  4 & 0.1236 &$1433$&$>14.78$& 1 & 0.1236 & $ 378$&$14.70$& 1\\
  & 12:19:40.27 & +06:40:41.2 & 0.1354 & E&  706 &  2.213 &  4 & 0.1350 &$ 452$&$15.12$& 3 & 0.1354 & $< 130$&$<14.10$& 9\\
  & 12:19:06.47 & +06:39:28.8 & 0.1810 & L &  632 &  2.391 &  3 & 0.1808 &$ 205$&$14.38$& 3 & 0.1810 & $<  80$&$<13.80$& 9\\
  & 12:19:04.05 & +06:41:35.6 & 0.1917 & U &  919 &  1.143 &  3 & 0.1917 &$<  45$&$<12.90$& 9 & 0.1917 & $<  40$&$<13.50$& 9\\
  & 12:19:20.39 & +06:36:57.6 & 0.2464 & E&  363 &  1.756 &  0 & 0.2464 &$<  45$&$<12.90$& 9 & 0.2464 & $<  40$&$<13.50$& 9\\
  & 12:19:34.84 & +06:40:55.8 & 0.2795 & L &  990 &  1.832 &  7 & 0.2786 &$  49$&$12.98$& 6 & 0.2795 & $<  60$&$<13.60$& 9\\
3C273 & 12:28:51.89 & +02:06:03.2 & 0.0902 & E&  443 &  1.105 &  0 & 0.0902 &$ 161$&$13.53$& 1 & 0.0902 & $  16$&$13.13$& 1\\
  & 12:29:24.10 & +02:08:12.5 & 0.1464 & L &  957 &  2.024 &  1 & 0.1466 &$ 414$&$13.94$& 3 & 0.1466 & $<   9$&$<12.87$& 4\\
PKS1302-102 & 13:05:20.22 & --10:36:30.4 & 0.0426 & L &  212 &  2.577 &  1 & 0.0422 &$ 410$&$14.83$& 3 & 0.0422 & $ 194$&$14.38$& 3\\
  & 13:05:14.84 & --10:40:00.4 & 0.0569 & E&  499 &  1.117 &  0 & 0.0569 &$< 350$&$<15.00$& 9 & 0.0569 & $<  40$&$<13.50$& 9\\
  & 13:05:41.01 & --10:26:42.5 & 0.0652 & E&  484 &  1.946 &  0 & 0.0647 &$ 256$&$14.06$& 3 & 0.0647 & $  69$&$13.70$& 3\\
  & 13:05:30.32 & --10:26:17.2 & 0.0718 & E&  540 &  2.399 &  1 & 0.0712 &$< 500$&$<16.00$& 9 & 0.0718 & $<  40$&$<13.50$& 9\\
  & 13:05:20.96 & --10:34:51.5 & 0.0939 & E&  330 &  5.381 &  9 & 0.0949 &$ 681$&$15.35$& 3 & 0.0948 & $  84$&$13.75$& 4\\
  & 13:05:28.64 & --10:38:14.9 & 0.1384 & E&  691 &  1.197 &  5 & 0.1393 &$<  35$&$<12.80$& 9 & \dots & \dots & \dots & \\
  & 13:05:40.17 & --10:36:54.2 & 0.1429 & E&  563 &  2.051 &  1 & 0.1429 &$<  40$&$<12.90$& 9 & 0.1429 & $<  40$&$<13.50$& 9\\
  & 13:05:34.00 & --10:39:59.6 & 0.1454 & L &  952 &  1.071 &  3 & 0.1453 &$ 890$&$15.29$& 3 & 0.1453 & $ 146$&$14.16$& 4\\
  & 13:05:31.29 & --10:35:42.1 & 0.1924 & E&  434 &  1.533 &  3 & 0.1916 &$ 401$&$15.01$& 3 & 0.1916 & $  80$&$13.93$& 1\\
FJ2155-0922 & 21:54:56.64 & --09:18:07.9 & 0.0517 & E&  251 &  1.427 &  0 & 0.0515 &$ 272$&$14.08$& 3 & 0.0514 & $<  47$&$<13.62$& 4\\
  & 21:55:22.71 & --09:23:51.8 & 0.0600 & E&  357 &  1.745 &  6 & 0.0600 &$<  30$&$<12.70$& 9 & 0.0600 & $< 120$&$<14.00$& 9\\
  & 21:55:17.31 & --09:17:51.8 & 0.0734 & E&  471 &  3.736 &  3 & 0.0736 &$ 232$&$14.64$& 3 & 0.0734 & $<  14$&$<13.06$& 4\\
  & 21:54:34.63 & --09:16:32.6 & 0.0785 & E&  741 &  1.574 & 21 & 0.0777 &$ 448$&$>14.29$& 1 & 0.0777 & $  46$&$13.61$& 1\\
  & 21:55:00.63 & --09:31:28.6 & 0.0831 & E&  792 &  2.979 &  1 & \dots & \dots & \dots & \dots &0.0831 & $<  50$&$<13.60$& 9\\
  & 21:55:06.53 & --09:23:25.2 & 0.1326 & L  &  212 &  1.626 &  2 & 0.1324 &$ 487$&$>14.30$& 1 & 0.1324 & $ 225$&$14.39$& 1\\
  & 21:54:54.91 & --09:23:30.8 & 0.1764 & E&  330 &  2.399 &  0 & 0.1765 &$ 479$&$>14.27$& 1 & 0.1765 & \ldots&$14.44$& 1\\
PKS2155-304 & 21:59:08.29 & --30:20:54.2 & 0.0454 & U &  423 &  1.142 &  2 & 0.0453 &$  87$&$13.40$& 3 & 0.0453 & $<   7$&$<12.76$& 4\\
  & 21:58:23.81 & --30:19:31.2 & 0.0541 & E&  543 &  2.053 &  0 & 0.0540 &$ 315$&$14.06$& 3 & 0.0540 & $  32$&$13.63$& 3\\
  & 21:58:40.81 & --30:19:27.1 & 0.0570 & E&  405 &  1.595 &  0 & 0.0566 &$ 477$&$14.48$& 3 & 0.0571 & $  44$&$13.57$& 3\\
\enddata
\tablenotetext{a}{Spectral type (E=Early-type, L=Late-type, U=unknown; see text for the quantitative definition).}
\tablenotetext{b}{Number of additional galaxies within 3\,Mpc of the sightline,  400\,\kms\ of this galaxy, and having $L>0.1L^*$.}
\tablerefs{1: \cite{tripp08}; 2: \cite{tc08a}; 3: \cite{ds08}; 4: \cite{dsr+06}; 5: \cite{pss04}; 6: \cite{cm09}; 9: This paper.}
\end{deluxetable*}

A closer inspection of Figure~\ref{fig:1Mpc} reveals that
the \lya\ detections for \lstar\ galaxies do not 
follow the trend of decreasing \wlya\ with increasing $\rho$.  
In fact, the \wlya\ values for these galaxies form two sets: the
galaxies with positive detections show $\mwlya > 100$\,m\AA\
(generally $\mwlya > 200$\,m\AA), yet approximately half of the
\lstar\ galaxies have no associated absorption to very sensitive limits
($\mwlya < 50$\,m\AA).  This phenomenon is highlighted in
Figure~\ref{fig:lstar_1mpc} where we plot \wlya\ vs.\ $\rho$ for
every \lstar\ galaxy within 1\,Mpc, showing the closest galaxy to the
sightline in cases where two or more \lstar\ galaxies lie within $\pm
400 \, \mkms$ of one another (Table~\ref{tab:lstar_1Mpc}). 
Again, the positive detections primarily result in $\mwlya >
200$\,m\AA\ and approximately half of the systems (without a fainter,
neighboring galaxy with smaller $\rho$) have a non-detection with
$\mwlya < 100$\,m\AA.
We find no clear differences in the \lya\ detection
rate or \wlya\ values for these galaxies
as a function of spectral type or redshift.
We hypothesize that the 
subset without a detection could be
sightlines penetrating
a hot (e.g.\ virialized) gas which has collisionally
ionized hydrogen to non-detectable levels.  The positive detections,
in contrast, would arise from gas that has not been shock-heated or
has since cooled.
This simple scenario, however, may not naturally explain the nearly
constant \wlya\ values for the positive detections.

To summarize the discussion thus far, our results (and previous work)
demonstrate that galaxies of all types are surrounded by a highly
ionized medium (which we term as an extended CGM)
with near unity covering fraction to \lya\ absorption to $\rho \approx
300$\,kpc.  Furthermore, galaxies of all types exhibit a declining
detection rate for \ion{H}{1} absorption with $\mwlya > 50$\,m\AA\ to $\rho \approx
1$\,Mpc.  A key question motivated by these results is what fraction
of the \lya\ forest arises from the CGM surrounding galaxies?  More bluntly, can the
virialized halos and extended CGM of galaxies account for the majority
of lines in the observed \lya\ forest?  These questions frame the
long-standing debate on whether \lya\ absorbers arise primarily from the
gaseous halos of galaxies or from the large-scale, overdense medium
(e.g.\ filaments) that encompass them.
To a large degree, the argument hinges on whether the structures
giving rise to \lya\ absorption have dimensions of several hundred kpc
or are greater than 1\,Mpc.  

It is evident from the results described above that a fraction of
\lya\ absorbers are closely
associated with galaxies. 
Indeed, some fraction of the \lya\ forest must arise from gas bound to
individual galaxies.
The majority of galaxy/absorber pairs in our sample, however, occur at impact
parameters that significantly exceed the presumed virial radius of the
galaxies. Indeed,
the measured \ion{H}{1} column densities are many orders of magnitude
smaller than the surface densities characteristic of star-forming
galaxies;  the gas cannot be related to canonical \ion{H}{1} disks
nor even the tidal features regularly observed in 21\,cm emission
observations \citep[e.g.][]{hibbard+01}.  
We conclude, therefore, that \lya\ absorption is generally
unrelated to the inner galaxy and/or the processes of galaxy
formation (e.g.\ winds, tidal stripping).  Instead, the
material we have associated with galaxies lies within their extended CGM.  In this case the
association need not be causal;  the gas may not be bound to the
galaxy and the two phenomena --- galaxies and
\ion{H}{1} absorbers --- may simply trace the same overdensities in the
universe.   

We draw further insight into the origin of the \lya\ forest
from the results presented in Figure~\ref{fig:lya_gal}.  
As has been emphasized previously
\citep{pss02,cpw+05,mj06}, the origin of \lya\ absorbers appears to be 
sensitive to the \ion{H}{1} column density.  
At low \nhi\ values, the fraction of `random' absorbers that may be
associated to a galaxy with $\rho < 300$\,kpc is small ($<20\%$
[15/82] for $\mnhi = 10^{13-14} \cm{-2}$).  This suggests
that the majority of weak \lya\ absorbers, which dominate the \lya\
forest by number, arise beyond the virialized halos and the CGM of
individual galaxies.  
Presumably, these absorbers arise in overdense regions of the \lowz\
universe which have not supported the formation of a luminous galaxy.
In contrast, the percentage of strong absorbers
($\mnhi > 10^{15} \cm{-2}$) with an associated galaxy at $\rho <
300$\,kpc is $80\%$ (8/10).  One is compelled, for this subset, to
physically associate the absorbers with the galaxies.
These inferences are supported by the \nhi\ dependence measured for
the clustering of \lya\ absorbers to \lowz\ galaxies
\citep{pss02,cpw+05,cm09,shone+10}. 
The much lower cross-correlation amplitude for systems with $\mnhi < 10^{14} \cm{-2}$
imply this gas is not physically associated to galaxies. The
high amplitude for $\mnhi > 10^{14.5} \cm{-2}$ absorbers, meanwhile, implies a
physical association with galaxies.  
These strong absorbers are sufficiently rare that one may associate each of them
with a galaxy. 
Indeed, the results in Figure~\ref{fig:lya_gal}
suggest that many of the strong absorbers arise within the 
virialized halos of individual galaxies.  Such environments
may be the only regions of the universe with sufficient overdensity to
give rise to \ion{H}{1} column densities exceeding $10^{15} \cm{-2}$.

\begin{figure*}[ht]
\includegraphics[height=6.5in,angle=90]{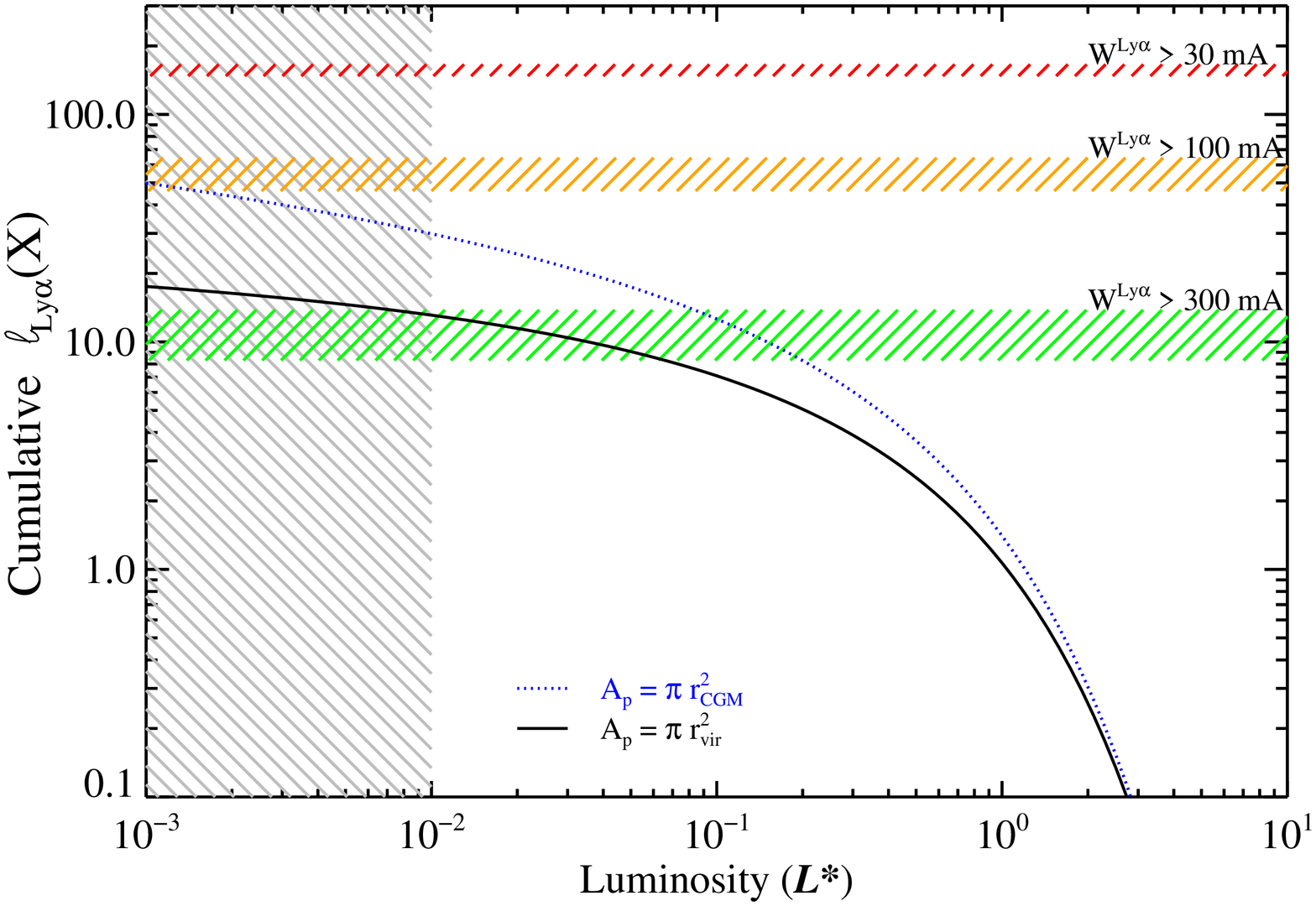}
\caption{The horizontal, hatched regions show measurements for the
  incidence of \lya\ at $z=0$ as a function of limiting equivalent
  width \citep{pss04}.  For strong absorbers ($\mwlya > 300$\,m\AA),
  we observe $\approx 10$ lines per unit pathlength (and redshift).
  The curves represent estimates for \lolx\ derived from the
  virialized halos of galaxies (black, solid) and the extended CGM
  surrounding galaxies (blue, dotted) integrated from a limiting
  luminosity.  The results indicate that the virialized halos of
  galaxies account only for the incidence of strong \lya\ lines
  observed in the \lowz\ IGM.  Even allowing for the extended CGM
  around each galaxy (with $\mrcgm = 300$\,kpc), the majority \lya\
  absorption in the \lowz\ IGM cannot arise from gas surrounding galaxies.
}
\label{fig:dndx_lya}
\end{figure*}

Another statistical approach to addressing these issues
is to compare the observed incidence of \lya\ absorption against that
predicted for gas surrounding galaxies based on our results and simplified
assumptions about the virial radii of galaxies
(Equation~\ref{eqn:rvir_pow}).  Specifically, we ask whether galaxy
halos occupy enough area to account for most \lya\ absorbers to a
given equivalent width limit.  To this end,
consider the incidence per unit pathlength $dX$ of \lya\
absorption \lolx, where 

\begin{equation}
dX = \frac{H_0}{H(z)}(1+z)^2 dz  \perd
\label{eqn:dX}
\end{equation}
At $z=0$, $dX= dz$ and \lox\ may be evaluated directly from the
observed incidence of absorbers per unit redshift\footnote{Commonly
  referred to as $n(z)$ or $dN/dz$.  Similarly \lox\ is often denoted
  $dN/dX$.}, \loz.  
The quantity \lox\ is defined to remain constant in time if the
product of the comoving number density $n_c$ of objects giving rise to
the absorption and the physical size $A_p$ of the sources remains
constant, i.e.\
\begin{equation}
\mlox = \frac{c}{H_0} \, n_c \, A_p \perd
\end{equation}
This evaluation assumes that 100\%\ of the area $A_p$ produces
detectable absorption.  One could include a factor in the
formalism that allows for a non-unity covering fraction.
Alternatively, one may consider $A_p$ to be the effective
area of the object to absorption.

In Figure~\ref{fig:dndx_lya}, hatched, colored bands show current
estimates for \lolx\ using results from $z=0$ surveys
of \lya\ absorbers \citep{pss04,wakker09}.  Each band 
corresponds to a limiting equivalent width, which for
low \wlya\ values translates directly to a limiting column
density.\footnote{$\mwlya = 50$\,m\AA\ roughly corresponds to $\mnhi = 10^{13} \cm{-2}$.}
Overplotted on Figure~\ref{fig:dndx_lya} is a black, solid curve that
shows the cumulative inferred contribution to \lolx\ of
virialized galactic halos as a
function of galaxy luminosity.  For this calculation, we have 
estimated $n_c$ from the galaxy luminosity function of
\cite{blanton03} taking $h=0.72$.  We have also assumed that $A_p^{\rm
  Ly\alpha} = \pi \mrvir^2$ with \rvir\ given by
equation~\ref{eqn:rvir_pow}. 
It is obvious that bright galaxies ($L>L^*$) can only contribute a small
fraction of the total incidence of \lya\ absorbers.  Integrating down to $L=0.01L^*$, the
approximate limit of the LCO/WFCCD survey, we infer $\ell^{\rm 
  r_{\rm vir}}_{\rm Ly\alpha} (X; L>0.01L^*) \approx 15$.  
\cite{tf05} recovered a similar estimate for galactic halos (in the
context of \ovi\ absorption) in an analysis that also accounted for
the clustering of galaxies.
The estimate lies below all of the \lolx\ estimates shown in
Figure~\ref{fig:dndx_lya}, except the strongest \lya\ lines
($\mwlya > 300$\,m\AA).  
We conclude, therefore, that
{\it the virialized halos of individual galaxies can only account for
the strongest \lya\ absorbers ($\mwlya > 300 {\rm \,m\AA}$) observed in the
\lowz\ IGM}.  On the other hand, the results presented in
Figures~\ref{fig:dwarfs}--\ref{fig:subls} indicate systems with $\rho <
\mrvir$ to $L>0.1L^*$ galaxies do exhibit a high covering fraction to
$\mwlya > 300$\,m\AA\ absorption.  One further concludes that {\it the
  majority of strong \lya\ absorbers must arise from gas in galactic
  halos}.

As emphasized throughout this section, 
galaxies are also surrounded by a CGM
with nearly unit covering fraction for $\mwlya > 50$\,m\AA\ 
to $\rho = 300$\,kpc (Figure~\ref{fig:1Mpc}b).  The dashed blue curve in
Figure~\ref{fig:dndx_lya} shows an estimate of the cumulative
incidence of \lya\ absorbers derived under the assumption that a
CGM with $r_{\rm CGM} = 300$\,kpc surrounds every galaxy with unit covering fraction,
i.e.\ $A_p^{\rm Ly\alpha} = \pi r_{\rm CGM}^2$, independent of
galaxy luminosity.  This increases the estimated \lolx\ value relative
to the estimate for virialized halos,
especially for galaxies with $r_{\rm vir} \ll 300$\,kpc.
Nevertheless, even the gas from the extended CGM of
galaxies cannot match the observed incidence of weak \lya\
absorbers.\footnote{Our estimate could be increased by a modest factor
  if we allowed that one galaxy may give rise to multiple \lya\
  absorbers.  On the other hand, \cite{bowen+02} have noted that \lya\
  `clusters' are likely associated with galaxy groups (i.e.\ multiple
  galaxies).}
We conclude that {\it the overwhelming majority of \lya\ lines detected
  from the IGM arise in structures located at distances beyond several
  hundred kpc from $L>0.01L^*$ galaxies.}
One must seek an additional origin for the majority of IGM absorption
then gas surrounding galaxies.

For nearly two decades, researchers have performed hydrodynamic
cosmological simulations of the IGM to simulate and study the \lya\
forest \citep[e.g.][]{mco+96,GnedHui98}.  This includes several analyses
of the low-$z$ IGM \citep{dhk+99,co99,rfb06,pjt+09}.  Regarding \lya\
absorption, the low-$z$ studies have focused primarily on the
distributions of \ion{H}{1} column density and Doppler parameter and
on associating the gas to various phases of the universe.  In their
recent publication, \cite{dave+10} have reported that the majority of
observable \lya\ absorption (systems with $\mnhi \approx 10^{13} $--$
10^{15} \cm{-2}$) is associated with diffuse, photoionized gas at
temperatures of $\sim 10^4$\,K, with density $n_{\rm H} \approx
10^{-4}$--$10^{-6} \cm{-3}$, and arising in structures with characteristic sizes of
several hundred kpc to 1\,Mpc.  
These structures, which trace large-scale overdensities in the \lowz\
universe, presumably also contain galaxies.  In their first paper,
\cite{dhk+99} examined the association of galaxies to the IGM and
found that the cosmological simulations reasonably reproduced the
observed high incidence of \lya\ absorption at small impact parameters
to galaxies.  They also roughly reproduced the observed relation
between equivalent width and impact parameter to $\rho \approx
1$\,Mpc.  The majority of these \lya\ absorbers arise in shocked or
diffuse gas surrounding the galaxies and the gas is not bound to them.  The
overall incidence of \lya, meanwhile, is dominated by gas at $\rho \gg
100$\,kpc from galaxies.  Their results appear,
at least qualitatively, to provide a reasonable
match to our new observations although we encourage future,
quantitative comparison.
We are further compelled, therefore, to support this model over ones
where virialized halos or the extended CGM associated with discrete 
galaxies dominate the observed IGM.

The cosmological simulations described above predict that the gas arises in a filamentary
network of overdensities, known as the ``cosmic web'', that permeates the universe.  
In this same
framework, galaxies form within the filaments and galaxy clusters
are produced at the nodes where filaments intersect.  The observed
distribution of galaxies in wide-field surveys generally
support this picture \citep{gott+09,bsc10}.
Inspired by this paradigm, we introduce a simple filament model to
interpret the trends observed in Figure~\ref{fig:1Mpc}. 
The model assumes that each galaxy in the
LCO/WFCCD survey resides at the center of a filamentary structure,
generally not at a node.  The filaments have
a finite width $w$ but are presumed to be infinitely 
long.\footnote{These filaments also have a finite depth whose dimension
  is irrelevant to the analysis.}
In this
simple scenario, the covering fraction to \lya\ absorption associated
with a galaxy is purely geometrical.  It is unity for $\rho < w/2$
and decreases at larger values.  Overplotted in
Figure~\ref{fig:1Mpc}b is the predicted covering fraction for this
simple model with $w = 400$\,kpc.  This model shows good agreement
with the observations despite
its simplicity.  It suggests that
the overdense structures hosting
$z\sim 0$ galaxies have characteristic dimension of several hundred kpc.
Indeed, \cite{dave+10} estimate that $w \sim 400$\,kpc corresponds to
absorbers with $\mnhi \approx 10^{13.3} \cm{-2}$, a characteristic
value for the gas analyzed here.
Lastly, if such filaments dominate the incidence of \lya\ in the
observed IGM then their filling factor of galaxies must be low
so that galaxies are only rarely associated with \lya\ absorption
(Figures~\ref{fig:lya_gal} and \ref{fig:dndx_lya}). We infer that only
rare `patches' of filaments contain $L > 0.01L^*$ galaxies.


\subsection{\ovi}
\label{sec:discuss_ovi}


Let us now consider \ovi\ gas and its relation to galaxies in the $z
\sim 0$ universe.  Aside from the \ion{H}{1} Lyman series, the most
commonly detected absorption lines of the \lowz\ IGM are the \ovi\
doublet.  This reflects the high number abundance of oxygen, the ease
of identifying a line doublet, and the predominance of regions within
the IGM that can produce the O$^{+5}$ ion.  With the launch of {\it
  HST} and its suite of UV spectrometers, \ovi\ has been surveyed from
$z \approx 0.2 $--$ 1$ \citep{bt96,trippetal00,tripp08,tc08a,ds08}.
Secondly, the {\it FUSE} spectrometer, which operated at $\lambda
\approx 910$--$1150$\AA\ for nearly a decade, enabled the search for
intergalactic \ovi\ down to $z=0$ \citep{pks0405_uv,dsr+06,wakker09}.  Together, these surveys
have discovered over 100 \ovi\ doublets to establish their incidence
as a function of equivalent width (and column density), to
characterize their relation to \ion{H}{1} gas, and to measure their
distribution of line widths.

Previous work has proposed a variety of origins
for this \ovi\ absorption.  On the smallest scales, the high incidence of
\ovi\ detections in the halo of our Galaxy \citep{sws+03}
implies a high filling
factor of such gas in the dark matter halos of $L \approx L^*$
galaxies.  
This assertion has now been confirmed, at least for blue and star-forming
\lstar\ galaxies at $z \sim 0.2$ (Tumlinson et al., in prep.). 
At the largest scales (i.e.\ intergalactic), 
cosmological simulations predict that
the WHIM comprises $\gtrsim 25\%$ of the baryons in the modern
universe, and some models also predict that this gas gives rise to a
significant fraction of the observed \ovi\ absorption \citep[e.g.][]{cf06}.  
In between these extremes, the intragroup medium
(revealed in some cases by X-ray emission) may have the density and
temperature required to yield significant \ovi\ absorption \citep{mulchaey96}.
Of course, each of these environments is likely to contribute, and each
could even dominate at different \ovi\ equivalent width.

A key approach to discriminating among these scenarios is to explore the
association between \ovi\ gas and galaxies.  
\cite{tf05} compared the incidence of \ovi\ absorption against
predictions for low-$z$ galaxies as characterized by SDSS and
concluded that bright galaxies ($L>L^*$) could not reproduce the
observed rate.  They suggested that fainter galaxies may explain the
observed incidence if dwarf galaxies were surrounded by an enriched
medium to $\approx 200$\,kpc.
\cite{stockeetal06} have correlated $L>0.1L^*$ galaxies with
9 \ovi\ absorbers in several fields and reported a median offset of
$\approx 180$\,\hvkpc.  They argued that \ovi\ gas is predominantly
associated with individual galaxies and speculated that galaxies with
$L< 0.1L^*$ dominate.  \cite{wakker09} surveyed 6 field galaxies at $z
\approx 0$ with $L \ge 0.1L^*$ and $\rho < 350$\,kpc and reported
\ovi\ detections for 2/3.  They also argued that gas in the extended
surroundings of galaxies contributes significantly to \ovi\ (and \lya)
absorption. 

The results presented in this paper offer new insight into the origin
of \ovi\ at modest to large equivalent widths
($\mwovi \gtrsim 30$\,m\AA).  
Our survey and analysis has focused on the properties of galaxy halos
and the extended CGM as related to \lya\ and \ovi\ absorption.
In contrast to \lya\ (as discussed in the previous subsection), the
association between galaxies and \ovi\ shows a strong dependence on
galaxy luminosity.
In $\S$~\ref{sec:step1} (Figure~\ref{fig:dwarfs}), we
showed that sightlines at impact parameters $\rho < 300$\,kpc to dwarf
galaxies ($L<0.1L^*$) rarely exhibit \ovi\ absorption.  Furthermore,
the few dwarfs with associated \ovi\ gas also had a neighboring $L>0.1L^*$
galaxy within $\rho = 200$\,kpc of the sightline.  
The implication is that the extended CGM of dwarf galaxies\footnote{ 
  Gas within the virialized halos of dwarf galaxies (i.e., at $\rho \lesssim
  50$\,kpc), however, was not well probed by our survey.}
does not contribute significantly to \ovi\ absorption with $\mwovi
\gtrsim 50$\,m\AA.  
For the brightest galaxies ($L>L^*$), we found a high incidence of \ovi\
detections for $\rho < 225$\,kpc but not a single detection
beyond (to sensitive limits; Figure~\ref{fig:lstar}, Table~\ref{tab:lstar}).  
In Figure~\ref{fig:lstar_1mpc}b, we have extended the search for \ovi\
associations with \lstar\ galaxies to 1\,Mpc.  Although there are a
few detections for $\rho > 300$\,kpc, the majority of these cases also
show an additional, fainter ($L<L^*$) galaxy located within $\rho = 300$\,kpc of the
sightline.   We conclude that the covering fraction of \ovi\ gas
around \lstar\ galaxies is
small for $\rho \gtrsim 250$\,kpc \citep[see also][]{wakker09}.

In contrast to the
faint and bright galaxies, the intermediate \subls\ population
exhibits a high incidence of \ovi\ at all impact parameters $\rho <
300$\,kpc (Figure~\ref{fig:subls}).  
This conclusion is independent of spectral type; we find associated
\ovi\ for \subls\ galaxies with early-type (presumed `red and dead') and
late-type (star-forming) spectra.
Combining these results, we
associate \ovi\ gas preferentially with the halos and CGM of \subls\
galaxies.  But this begs the obvious question, are these galaxies
sufficiently common that they trace the majority of \ovi\ in the
universe?  Or does \ovi\ also arise from an additional reservoir, as
we found for weak \lya\ absorbers?

\begin{figure*}
\includegraphics[height=6.5in,angle=90]{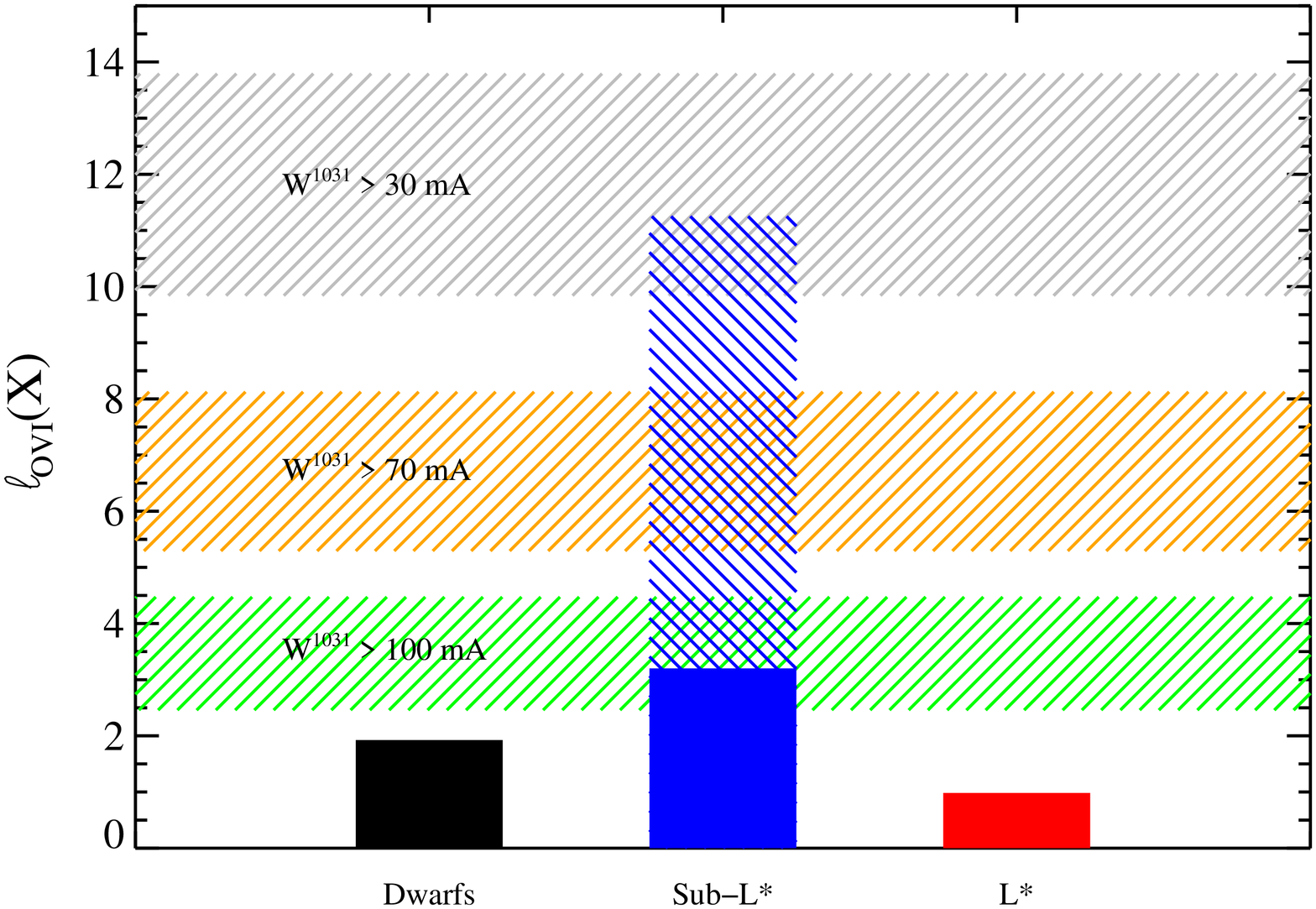}
\caption{Horizontal bands show the observed incidence of \ovi\
  absorption per absorption length \loox, as a function of limiting
  equivalent width \citep{tripp08}.   The solid vertical bars indicate
  estimates of the incidence of \ovi\ absorption for gas in the
  virialized halos of dwarf (black; $0.01 L^* < L < 0.1 L^*$), \subls\
  (blue; $0.1L^* < L < L^*$), and $L>L^*$ (red) galaxies.  For these 
  estimates we have assumed virial radii of $\mrvir = 100,160$ and
  250\,kpc respectively, and that the halos have a covering fraction of 100\%\
  to \ovi.  Under this parameterization, virialized halos of galaxies
  only account for stronger \ovi\ systems.  The
  hatched, blue bar gives the estimate of \loox\ for gas in the
  extended CGM of \subls\ galaxies assuming a 100\%\ covering fraction
  to $\rho = r_{\rm CGM} = 300$\,kpc, as suggested by our LCO/WFCCD
  survey
  (Figure~\ref{fig:subls}).  The excellent correspondence with the
  observed \loox\ value for $\mwovi > 30$\,m\AA\ indicates that 
  the remainder of \ovi\ systems arise in this extended CGM.  
  We conclude that the majority of \ovi\
  gas observed in the \lowz\ IGM arises in the 
  diffuse medium surrounding individual galaxies (predominantly $L
  \approx 0.3 L^*$), and that it rarely originates in the 
  WHIM predicted by cosmological simulations.
}
\label{fig:dndx_ovi}
\end{figure*}

In Figure~\ref{fig:dndx_ovi}, we plot the incidence per unit pathlength
$dX$ of \ovi\ absorption, \loox,
estimated from the incidence per unit redshift
\looz\ measurements of \cite{tripp08} for several equivalent width limits.
Formally, we have converted their estimates of the incidence per unit
redshift \loz\ measured at $z \approx 0.2$ to \lox\
(Equation~\ref{eqn:dX}), which
implies a downward correction of $\approx 30\%$. 
The width of the hatched regions indicate the $1\sigma$ uncertainties
reported by \cite{tripp08} and we caution that systematic bias may
have led to a modest overestimate in these \loox\ evaluations
\citep{tc08b}. 
Overlaid on the observational estimates of \loox, we plot the predicted
incidences of \ovi\ from each of our galaxy subsets.  
The solid bars
show the estimated contribution from gas within the virialized halos
of each galaxy sample.  In each case, we
have estimated $n_c$ from the galaxy luminosity function\footnote{  
 For the dwarf galaxy calculation,
  we limit the integration to $0.01 L^* < L < 0.1 L^*$.} 
of \cite{blanton03} taking $h=0.72$.  
For the effective cross-section of halos, we have assumed $A_p^{\rm OVI} = \pi
\mrvir^2$ with $\mrvir = 100, 160, 250$\,kpc  
for the dwarf, \subls, and $L^*$ populations, respectively.   
These values are somewhat smaller than the value one would derive by
averaging Equation~\ref{eqn:rvir_pow} over the luminosity function as
our estimates for $A_p^{\rm OVI}$ include a modest correction for the
covering fractions to \ovi\ absorption.

None of the galaxy populations (on their own) can account for the incidence of the commonly
observed \ovi\ absorbers if one restricts to virialized halos. 
For the \lstar\ galaxies, the estimated incidence is $\approx 10\%$ of
the total, in line with the model predictions by \cite{tf05}.
Furthermore, dwarf galaxies may give rise to as much as 20--30$\%$ of
the total incidence but we caution that the estimate shown in
Figure~\ref{fig:dndx_ovi} assumes a higher covering fraction to \ovi\
for dwarfs than supported by our observations
(Figure~\ref{fig:dwarfs}).
Together, galactic halos can account for the incidence of
very strong \ovi\ absorption ($\mwovi > 100$\,m\AA) and possibly
moderate \ovi\ absorbers ($\mwovi \approx 70$\,m\AA) .  
We conclude, however, that {\it gas in the virialized halos of $z\sim
  0$ galaxies with $L > 0.01L^*$ cannot account for the majority of 
  observed \ovi\ systems} ($\mwovi > 30$\,m\AA).
A similar inference was drawn by \cite{tf05}.

The above conclusion is also supported by the results presented in
Figure~\ref{fig:ovi_gal}, 
where we found that very few \ovi\ absorbers are coincident with 
the virialized halo of a galaxy.
We further emphasize that the $z \sim 0$ luminosity function is too
shallow $(\Phi \propto L^{-1})$ for fainter galaxies 
to qualitatively modify this conclusion.\footnote{Although there remains debate on
  the precise slope of the faint-end, one would require a much steeper
  slope than currently estimated for faint galaxies to significantly
  contribute.}  
On the other hand, one may attribute the 
strongest \ovi\ absorbers ($\mwovi > 100$\,m\AA) to virialized halos
and, by inference, galactic-scale processes.
In fact, one must conclude from the results presented in
Figures~\ref{fig:dwarfs}-\ref{fig:subls}, \ref{fig:ovi_gal}, and
\ref{fig:dndx_ovi}
that {\it the majority of strong \ovi\ systems 
arise from with the halos of individual galaxies.}

With virialized halos ruled out as the dominant origin of \ovi\ gas
(by frequency of absorption not necessarily by mass),
we now consider the extended CGM surrounding $z \sim 0$ galaxies.  As
emphasized above, we find a significant covering fraction to $\rho =
300$\,kpc for the \subls\ galaxies (Figure~\ref{fig:subls}).
Overlaid on the \loox\ estimate for virialized halos of the \subls\ galaxies is a hatched
bar that represents the contribution to \loox\ from an extended
CGM surrounding each \subls\ galaxy.  
This hatched bar assumes 
$A_p^{\rm OVI} = \pi \mrcgm^2$ with $\mrcgm = 300$\,kpc and a 100\%\ covering
fraction for $\mwovi \ge 30$\,m\AA.  This amounts to a several times higher
incidence of \ovi\ absorption. 
Remarkably, we find that the extended CGM of \subls\ galaxies
reproduces 
the incidence of all \ovi\ absorbers with $\mwovi \gtrsim 30$\,m\AA!
Of course, these results follow from the close association of galaxies
to \ovi\ absorbers (Figure~\ref{fig:ovi_gal}) and the high covering
fraction of \ovi\ gas surrounding \subls\ galaxies
(Figure~\ref{fig:subls}).
While one could
reduce this estimate by a factor of order unity (e.g.\ by assuming a smaller
covering fraction), {\it we conclude that the extended CGM of \subls\
galaxies is the primary reservoir of \ovi\ gas observed in the $z \sim
0$ IGM}.  This final conclusion, which draws a distinction between
virialized halos of \subls\ galaxies and their extended CGM, is
sensitive to the precise definition of a galaxy halo and its radius.
We encourage additional theoretical exploration into the nature and
extent of virialized gas in the halos of subluminous, \lowz\ galaxies.

Let us now consider the implications of an extended CGM 
that gives rise to significant \ovi\ absorption surrounding
\subls\ galaxies.  First, we note that this CGM 
may include neighboring galaxies, the intragroup
medium, and the diffuse medium that envelops the galaxies (e.g.\ filaments).  
Regarding neighboring galaxies,\footnote{Also, be definition the volume
  beyond \rvir\ must have a lower density than the medium within it.} 
we have already demonstrated that 
virialized halos contribute only modestly to the incidence of
\ovi.  By
the same argument, they cannot dominate the detections associated with
the extended CGM of \subls\ galaxies.  Regarding the intragroup
medium, only a subset of \subls\ galaxies exhibit galaxy overdensities
suggestive of a group (Table~\ref{tab:subls}).  We infer, therefore,
that this extended CGM is best described by a diffuse, modestly
overdense ($\delta \gtrsim 10$) medium.  
In the IGM paradigm of undulating
Gunn-Peterson absorption, such overdensities
naturally give rise to significant \ion{H}{1} absorption.  Indeed,
as discussed above, \lya\ absorption is a ubiquitous phenomenon for
galaxies of all luminosity, and every \ovi\ absorber detected to date
shows corresponding \lya\ absorption \citep{tc08a}.

Although one may naturally expect \ion{H}{1} absorption from an
overdense region of the universe, the detection of \ovi\ further
requires a chemically enriched medium and the physical conditions that
produce the O$^{+5}$ ion.  Regarding the former, our results require
that the CGM of \subls\ has been polluted by metals on scales of
several hundred kpc.  Although many of the \subls\ galaxies in our
sample show the spectral signatures of ongoing star-formation, these
are not star-bursting systems, and they are unlikely to currently be driving
galactic-scale winds to such large distances.  We infer, therefore,
that this CGM was previously enriched.  Current estimates for
the metallicites of the \ovi\ absorbers suggest values ranging from
0.01 to nearly solar abundance with a median of $\approx 0.1$ solar
\citep[e.g.][]{pks0405_uv,cpc+08,tc08b,ds08}.  
This enrichment level exceeds predictions from the first stars
\citep[e.g.][]{wtn+10} indicating 
a subsequent phase of star-formation is required.  The medium is too
diffuse to form metals 
{\it in situ}, and therefore, these must have been
transported by distances of 100\,kpc or more.
The proximity of the \subls\ galaxy suggests that the oxygen was
produced in that system, yet this is purely speculative.
We also note, following the mass estimate from
$\S$~\ref{sec:discuss_lya} (Equation~\ref{eqn:mass_CGM}), that a 0.1
solar metallicity implies a metal mass equivalent to $\approx 3 \sci{9} \msun$
of solar metallicity stars.  This could exceed the total mass in
metals within a typical \subls\ galaxy and its satellites, which would
require mass outflow rates from galactic winds that match or exceed
the star-formation rate \citep[e.g.][]{wcp+09}.
Comprehensive modeling of the chemical enrichment of the IGM that
includes galaxy formation and metal transport are required to further explore this topic
\citep[e.g.][]{co06,gcf+08,od08,od09}.

In addition to containing oxygen, this CGM must have the
appropriate density and/or temperature for a significant fraction of
oxygen to exist as O$^{+5}$.  
One process to produce O$^{+5}$ is via collisional ionization (CI) which
requires temperatures on the order of $T \sim 10^5 $--$ 10^6$\,K.
The other obvious mechanism to produce
O$^{+5}$ is with photoionization, which requires photons with
energies $h\nu > 8$\,Ryd.  Such hard photons are only produced by AGN, 
and the extragalactic UV background (EUVB) is the obvious source
of such radiation when one is far from active galaxies. 

Consider first several inferences one may draw from our association of
\ovi\ to the extended CGM of \subls\ galaxies.
If the extended CGM (almost by definition) 
lies beyond the virialized
radius of these galaxies, one may not expect a collisionally ionized
medium. Furthermore, the majority of these 
\subls\ galaxies are not obviously located within a larger dark
matter halo (e.g.\ a group or cluster).  
Therefore, this gas may not have been virialized to $T \gg
10^4$\,K.
One would require instead gravitationally-induced shock-heating from larger scales.
Indeed, the collapse of large-scale density `waves' is predicted to drive the production of
the WHIM in cosmological simulations \citep[e.g.][]{co99,daveetal01} and may, in principle, heat the
CGM of \subls\ galaxies.  Although this is plausible, these waves are generally
predicted to `crash' on larger scales (many Mpc) than the several
hundred kpc scales of the CGM.    
Indeed, \cite{gcf+08} analyzed the average impact parameter to
galaxies of \ovi\ gas arising in the WHIM.
They found in their simulations that only $\approx 1\%$ of \ovi\
detections with $\mwovi \ge 50$\,m\AA\ would have a $L>0.01L^*$ galaxy
within 300\,kpc.
This prediction is strongly ruled out by our observations \citep[see
also][]{wakker09}. By inference, we conclude that {\it \ovi\ gas
observed in the \lowz\ IGM does not typically trace a collisionally 
ionized WHIM}.

Empirically,
the ionization mechanism for observed \ovi\ absorbers remains a matter
of great debate.  We support (and have contributed to) arguments that
conclude most \ovi\ systems are not
collisionally ionized \citep{pks0405_uv,tripp08,tc08b}.  The evidence
includes the close alignment of \ion{H}{1} gas with \ovi, the
narrowness of the
coincident \lya\ absorption (and of \ovi\ itself), and the very frequent detection
of coincident \ion{C}{3} absorption.
On the other hand, see \cite{fox11} for a set of arguments against
\ovi\ arising in predominantly photoionized gas.  Given the results
presented above, we may at least
consider whether a photoionized, extended
CGM of \subls\ galaxies may naturally support \ovi\ absorption.

Under the assumptions that the gas is optically thin to ionizing
radiation and that the UV radiation field is dominated by a typical
AGN spectrum ($f_\nu \propto \nu^{-1.6}$), standard photoionization
calculations indicate that an ionization parameter\footnote{$U\equiv
  \Phi/c n_{\rm H}$, where $\Phi$ is the flux of ionizing photons.}  
$U \gtrsim 10^{-1}$ is required to produce a significant fraction of
O$^{+5}$ ions.  
To achieve $U \gtrsim 10^{-1}$ with current estimates for the EUVB
\citep[e.g. CUBA, which predicts $J_\nu \approx 10^{-25} \, {\rm erg
  \, s^{-1} \, Hz^{-1} \, \cm{-2}}$ at 8\,Ryd;][]{hm96}, this
demands a volume density $\mnh \lesssim 10^{-5} \cm{-3}$.  This value
is roughly $50\times$ the mean baryonic density at $z \sim 0$
suggesting an overdensity $\delta \lesssim 50$. 
These same photoionization calculations imply a size for the structure of $d \sim
300$\,kpc ($\mnhi/10^{14} \cm{-2}$).  
The constraints on $n_H$ and $d$ give plausible values for the
extended CGM of \subls\ galaxies.  They imply a gas that is overdense
yet not virialized ($\delta \ll 200$).  Furthermore, the inferred size
of the structures coincides with the dimensions inferred from our
analysis.  We conclude that the gas in the extended CGM could be
predominantly photoionized material giving rise to \ovi\ absorption.

We have developed the following picture for \ovi\ absorption in the
\lowz\ IGM: the gas has density $n_H \approx 10^{-5} \cm{-3}$, is
predominantly photoionized, and is 
primarily located in the extended CGM of \subls\ galaxies.  The oxygen
was produced in a previous episode of star-formation (perhaps by the
observed \subls\ galaxy) and has been transported to the CGM by one or
more processes.  Remarkably, this scenario is qualitatively consistent
with the model for \ovi\ described by \cite{od09}.  In their
cosmological simulations, oxygen is produced at earlier times (often
$z>1$) in a star-bursting galaxy whose galactic-scale wind transports
the metals to the surrounding, overdense IGM.  This gas is observed
today primarily as photoionized \ovi\ and is coupled to significant \ion{H}{1}
absorption mainly because both exist in overdense regions.

\cite{od09} studied the equivalent width and Doppler parameter
distribution of \ovi\ gas and found their model gives reasonably good
agreement to the observed distribution  
\footnote{These authors had to include turbulent
  broadening, added in post-processing, to explain the observed
  linewidths of \ovi.}.
These authors also performed
a qualitative analysis on the connection between \ovi\ absorption and
galaxies (their Figures~14 and 15).  They found \ovi\ gas is generally
located at several virial radii ($\sim 100$--$300$\,kpc) from galaxies
with masses  $M_{\rm gal} \sim 10^{9.5-10} \msun$ ($\sim 0.03 $--$0.1
M^*$).  Although these separations are consistent with our results, 
the typical galaxy mass may be lower than that of the average \subls\ galaxy.  We
encourage further analysis to consider the covering fraction to \ovi\
absorption as a function of galaxy luminosity
\citep[e.g.][]{gcf+08}.  The data also permits one to examine trends
between the galactic mass and environment with the observed equivalent
widths of absorption.

Before concluding, 
it is worth speculating
on why the extended CGM for \lstar\ and dwarf galaxies would have a lower
covering fraction than that observed for \subls\ galaxies.  
Regarding dwarf galaxies, there are several plausible explanations.
First, these galaxies generally show lower \nhi\ values on average
suggesting lower total gas columns.  Second, it is possible that the
CGM of dwarf galaxies is chemically poor in comparison to brighter
galaxies.\footnote{An ongoing Cycle~18 {\it HST}/COS program will
  assess the chemical enrichment of gas associated with dwarf galaxies
  (PI: Tumlinson; 1224).}
Lastly, the physical conditions of the gas (temperature, density) in
the CGM surrounding dwarf galaxies may not support the O$^{+5}$ ion,
e.g.\ the gas may have too low volume density.
Together, these effects could reduce the associated
\ovi\ absorption to equivalent widths below typical detection limits.

For the \lstar\ galaxies, however, we observe \ion{H}{1} column
densities at least as large as those observed for the \subls\ galaxies
(Table~\ref{tab:gastat}).  One may also expect the gas to have higher
metallicity.\footnote{Although, one could also speculate that the larger
potential well of \lstar\ galaxies prevents the transport of metals
beyond their virialized halos \citep[e.g.][]{od08}.  
This hypothesis, however, is apparently
not supported by current research on galactic-scale winds
\citep[e.g.][]{wcp+09,rwk+10}.}  
If the lower incidence of \ovi\
absorption is unrelated to differences in chemical enrichment, then the
result must relate to the physical conditions of the gas, i.e., the
extended CGM of \lstar\ galaxies has a density and/or temperature that
is not favorable for O$^{+5}$ ions.  Under the
assumption of photoionization by the EUVB, the gas must have $\mnh
\approx 10^{-5} \cm{-3}$ for its ionization potential to favor
O$^{+5}$ ions.  If the CGM of \lstar\ galaxies is $\approx 10\times$ higher
than that of \subls\ galaxies at $\rho \sim 200$\,kpc, then this could
explain the preponderance of non-detections.  In this scenario, one
would predict the detection of metals at $\rho \sim 200$\,kpc from
\lstar\ galaxies in lower ionization states (e.g.\ \ion{C}{4},
\ion{C}{3}).  
One might then expect a higher incidence of detections at somewhat
larger impact parameters,\footnote{Also, low column densities at all
  impact parameters if one assumes spherical symmetry.}
but this is not observed (Figure~\ref{fig:lstar_1mpc}b).   
As a final alternative, it is possible that gas at
$\rho \sim 250$\,kpc from an \lstar\ galaxy has been shock-heated to
$T \gg 10^5$\,K. Indeed, these galaxies are more frequently members
of a group, i.e.\ embedded within a larger, virialized halo.

\section{Summary}
\label{sec:summary}

With this paper we have examined the associations of galaxies to \lya\
and \ovi\ absorption in the \lowz\ IGM to explore the origin of this
gas.  Specifically, we have correlated galaxies from our LCO/WFCCD
galaxy survey \citep{ovi_paper4} and IGM absorption from published
line-lists of {\it HST} and {\it FUSE} UV quasar spectra.
These two datasets are essentially independent of one another,
although we have supplemented the IGM linelists with our own analysis
of the quasar sightlines.
The following bullets summarize our main findings:

\begin{itemize}
\item Galaxies of all luminosity ($L>0.01L^*$) and spectral-type
  show strong, associated \lya\ absorption to impact parameter $\rho \approx 300$\,kpc
  with a very high covering fraction ($\approx 90\%$). The strongest
  \ion{H}{1} absorbers ($\mwlya \gtrsim 1$\AA) are preferentially
  associated with brighter galaxies ($L>0.1L^*$).  
\item We estimate a baryonic mass for this extended circumgalactic medium (CGM) of $M_{\rm CGM}
  \approx 3 \sci{10} M_\odot \; (\mncgm/10^{19} \cm{-2}) (\mrcgm/300
  \, \rm kpc)^2$, having adopted a constant \ion{H}{1} column density
  $\mncgm$.
\item  Galaxies with luminosities $L>0.1L^*$ exhibit a high covering fraction
  ($>80\%$) 
  for significant \ovi\ absorption to $\rho = 200$\,kpc and to 300\,kpc for
  \subls\ galaxies.  Dwarf galaxies ($L < 0.1L^*$) exhibit a low
  covering fraction for $\rho > 50$\,kpc.
\item Despite these high covering fractions of \lya\ and \ovi\
  absorption, there are examples of non-detections to very sensitive
  limits, even at very low impact parameters $(\rho \lesssim 50$\,kpc).
\item We observe a declining covering fraction and median equivalent
  width for \lya\ to $\rho = 1$\,Mpc.  The equivalent widths of the positive
  detections at $\rho > 100$\,kpc may be described by a
  power-law, $\mwlya(\rho) = 3.3 {\rm \AA} \, (\rho/1\,\rm
  kpc)^{-0.43}$, but with great scatter.
\item To $\rho = 1$\,Mpc, \lstar\ galaxies show two distinct distributions of
  \lya\ absorption: (i) a set of positive detections with $\mwlya\ >
  200$\,m\AA; and (ii) a set of non-detections with $\mwlya <
  50$\,m\AA.
\item The detection rate of \lya\ absorption with impact parameter may
  be described by galaxies within filaments having characteristic
  widths $w \sim 400$\,kpc.
\item Few, if any, of the weak \lya\ absorbers ($\mwlya < 100$\,m\AA)
  from the \lowz\ IGM arise in the
  virialized halos of $z \sim 0$ galaxies, or their surrounding CGM.   
  The majority of strong \lya\ absorbers ($\mwlya
  > 300$\,m\AA), however, does arise in these environments.
\item The strongest \ovi\ absorbers ($\mwovi > 100$\,m\AA; $\mnovi >
  10^{14} \cm{-2}$) arise preferentially in the
  galactic halos of $L>0.01L^*$ galaxies.  Weaker \ovi\ absorbers are
  associated with the extended CGM of \subls\ galaxies.
\item Current predictions for models
  where \ovi\ gas arises in a collisionally ionized WHIM are ruled
  out at high confidence.  We suggest that \ovi\ gas is primarily
  associated with a photoionized CGM with $n_H \approx 10^{-5}
  \cm{-3}$ and typical dimension $d \sim 300$\,kpc.
\end{itemize}


\acknowledgments

J. X. P. is partially supported by an NSF CAREER grant (AST--0548180).
We acknowledge valuable comments and criticism of this work from
B. Oppenheimer, G. Worseck, J. Tumlinson, and A. Fox.





\end{document}